%% file: main.tex
\documentclass[sigconf]{acmart}

\AtBeginDocument{%
  }

\usepackage{amsmath,amsfonts}
\usepackage{graphicx}
\usepackage{textcomp}
\usepackage{xcolor}
\usepackage{tcolorbox}
\usepackage{listings}
\usepackage{proof}
\usepackage{amsthm}
\usepackage{multirow}
\usepackage{threeparttable}
\usepackage{booktabs}
\usepackage{listings}
\usepackage[ligature, inference]{semantic}
\usepackage{caption}
\usepackage{subcaption}
\usepackage{hyperref}
\usepackage{diagbox}
\usepackage{wrapfig}
\usepackage{tikz}
\usepackage{makecell}
\usepackage{enumitem}
\usepackage{todonotes}
\usepackage{algorithm}
\usepackage{algorithmicx}
\usepackage{algpseudocode}

\input{macros}

\setlist[itemize]{leftmargin=*, itemsep=0pt, topsep=5pt}

\begin{document}


\title{\toolname{}: Constraint-based Fuzz Driver Generation with Dual Scheduling}

\author{Yan Li}
\authornote{*First author}
\email{liyann@mail.ustc.edu.cn}
\affiliation{%
  \institution{University of Science and Technology of China}
  \city{Hefei}
  \country{China}
}

\author{Wenzhang Yang}
\email{wzhyang@iaii.ac.cn}
\affiliation{%
  \institution{Institute of AI for Industries}
  \city{Nanjing}
  \country{China}
}

\author{Yuekun Wang}
\email{ykwang@smu.edu.sg}
\affiliation{%
  \institution{Singapore Management University}
  \city{Singapore}
  \country{Singapore}
}

\author{Jian Gao}
\email{gaojian0074@163.com}
\affiliation{%
  \institution{Central University of Finance and Economics}
  \city{Beijing}
  \country{China}
}

\author{Shaohua Wang}
\email{davidshwang@ieee.org}
\affiliation{%
  \institution{Central University of Finance and Economics}
  \city{Beijing}
  \country{China}
}

\author{Yinxing Xue}
\authornote{†Corresponding author}
\email{yxxue@iaii.ac.cn}
\affiliation{%
  \institution{Institute of AI for Industries}
  \city{Nanjing}
  \country{China}
}

\author{Lijun Zhang}
\email{zhanglj@ios.ac.cn}
\affiliation{%
  \institution{Institute of Software, Chinese Academy of Sciences}
  \city{Beijing}
  \country{China}
}

\begin{abstract}

Fuzzing a library requires experts to understand the library usage well and craft high-quality fuzz drivers, which is tricky and tedious. 
Therefore, many techniques have been proposed to automatically generate fuzz drivers. 
However, they fail to generate rational fuzz drivers due to the lack of adherence to proper library usage conventions, such as ensuring a resource is closed after being opened.
To make things worse, existing library fuzzing techniques unconditionally execute each driver, resulting in numerous irrational drivers that waste computational resources while contributing little coverage and generating false positive bug reports.

To tackle these challenges, we propose a novel automatic library fuzzing technique, \toolname{}, an LLM-based library fuzzing technique. 
It leverages LLMs to understand rational usage of libraries and extract API combination constraints. To optimize computational resource utilization, a dual scheduling framework is implemented to efficiently manage API combinations and fuzz drivers.
The framework models driver generation and the corresponding fuzzing campaign as an online optimization problem. Within the scheduling loop, multiple API combinations are selected to generate fuzz drivers, while simultaneously, various optimized fuzz drivers are scheduled for execution or suspension.

We implemented \toolname{} and evaluated it in 33 real-world libraries. Compared to baseline approaches, \toolname{} significantly reduces computational overhead and outperforms \utopia{} on 16 out of 21 libraries. It achieves 1.62$\times$, 1.50$\times$, and 1.89$\times$ higher overall coverage than the state-of-the-art techniques \ckgfuzzer{}, \promptfuzz{}, and the handcrafted project \ossfuzz{}, respectively.
In addition, \toolname{} discovered 33 previously unknown bugs in these well-tested libraries, 3 of which have been assigned CVEs.

\end{abstract}

\maketitle

\input{sections/0_intro}
\input{figures/valid_but_irrational}
\input{figures/overview}
\input{sections/1_background}

\input{sections/2_0_design}

\input{sections/5_0_eval}
\input{sections/6_0_related_work}
\input{sections/7_0_discussion_limitation}

\input{sections/8_0_conclusion}
\bibliographystyle{ACM-Reference-Format}
\bibliography{sample}

\appendix
\input{sections/11_appendix}

\end{document}

%% file: macros.tex
\newcommand{\toolnameplain}{Scheduzz}

\newcommand{\toolname}{\textsc{\toolnameplain{}}}

\newcommand{\groupschedule}{\textsc{GS}}
\newcommand{\driverschedule}{\textsc{DS}}
\newcommand{\groupgen}{\textsc{GG}}
\newcommand{\utopia}{\textsc{Utopia}}
\newcommand{\promptfuzz}{\textsc{PromptFuzz}}
\newcommand{\ossfuzz}{\textsc{OSS-Fuzz}}
\newcommand{\ckgfuzzer}{\textsc{CKGFuzzer}}

\newcommand*\circled[1]{\tikz[baseline=(char.base)]{
            \node[shape=circle,draw,inner sep=0.5pt] (char) {#1};}}

\newcommand{\challengeone}{{\circled{1}}}
\newcommand{\challengetwo}{{\circled{2}}}

\newcommand{\boldparagraph}[1]{\vspace*{0.5ex}\noindent{\textbf{#1}}}


\newcommand{\inlinecpp}[1]{\lstinline[language=C,basicstyle=\ttfamily\footnotesize]!#1!}

%% file: sections/0_intro.tex
\section{Introduction}
Fuzz testing is a well-known technique for automatically detecting program bugs and vulnerabilities. It has been widely used and has shown its effectiveness in practice. For example, OSS-Fuzz~\cite{ossfuzz} is a continuous fuzzing project for over 1,000 open-source software projects, and it identifies and fixes over 10,000 vulnerabilities and 36,000 bugs by August 2023. The basic idea of fuzzing is straightforward: a fuzzer automatically generates a bunch of program inputs for a fuzzing subject and identifies a defect when an error occurs.

However, appropriate fuzzing subjects are difficult to obtain in library fuzzing. Many exported functions serve as development APIs in a library, necessitating the use of fuzz drivers to combine these APIs logically and construct a valid executable program to accept fuzzing inputs. Rigorous fuzzing of a library requires numerous fuzz drivers, and manually constructing them is tedious and tricky. Unfortunately, the automatic generation of fuzz drivers still faces two primary challenges. The \textbf{challenge}\challengeone{} is generating not only syntactically and semantically valid API calling sequences but also rational ones. For instance, in the compression library \textit{snappy}~\cite{snappy}, the function \textit{RawUncompress} requires the uncompressed length from the function \textit{GetUncompressedLength}. Thus, \textit{GetUncompressedLength} must be called before \textit{RawUncompress} in a fuzz driver. 
A \textit{snappy} fuzz driver is considered irrational if it invokes \textit{RawUncompress} without including \textit{GetUncompressedLength}.
Existing work\cite{green2022graphfuzz, jeong2023utopia} employs extra information to guide API combinations or randomly combines APIs and filters parts of them without understanding the library's usage, leading to these approaches being inefficient and costly.
The \textbf{challenge}\challengetwo{} lies in the explosion of the number of API combinations and the generated drivers.
We refer to an API combination as an API group, and the space complexity of an API group is $O(2^N)$, where $N$ is the number of exported functions in a library. Furthermore, each API group can generate a corresponding fuzz driver. Consequently, it is impossible to generate drivers for all API groups and execute all of them due to limited computational resources in the real world. 

In this paper, we present \toolname{}, a tool designed to automatically understand API combination conventions and then generate and schedule fuzz drivers to effectively fuzz a library. Overall, \toolname{} extracts the type information of each function as explicit constraints and employs a Large Language Model (LLM) to infer implicit constraints to model the relationship between two APIs. Consequently, a solver determines the rational API group within these two types of constraints. Subsequently, \toolname{} initiates its dual schedulers. The two schedulers work asynchronously to select optimal API groups and fuzz drivers for fuzzing the target library in a loop.

To address \textbf{challenge}\challengeone{}, we propose explicit and implicit constraints to model valid and rational API combinations. Specifically, the group generation component (\groupgen{}) of \toolname{} identifies the combination of two APIs as explicit constraints if they share parameter types in their input and return values. These explicit constraints indicate possible control flows between two APIs. Moreover, implicit constraints encode whether two APIs should be composed from the developer's perspective. To obtain the implicit constraints from a library, \groupgen{} employs an LLM to understand the code and comments. Furthermore, we leverage the logic programming language Prolog~\cite{DBLP:journals/corr/abs-1011-5332} to solve both constraints and obtain a set of valid and rational API groups. 
Each group consists of multiple APIs that can be combined.

To tackle \textbf{challenge}\challengetwo{}, we devise a dual scheduling framework for API groups and fuzz drivers. The two dual schedulers run asynchronously and incorporate fuzzing execution feedback. In each round, the \textit{Group Scheduler} (\groupschedule{}) selects a batch of API groups using four metrics and constructs a prompt to generate fuzz drivers with an LLM. Consequently, some fuzz drivers are added to a fuzz driver pool if they pass driver filters. Otherwise, they are sent back for re-generation. Meanwhile, the \textit{Driver Scheduler} (\driverschedule{}) selects drivers and begins executing them with a CPU time slice. After each slice is finished, \driverschedule{} pauses the running fuzzers and places them back into the fuzz driver pool. The execution information is collected as feedback for \groupschedule{} and \driverschedule{}.

We evaluated \toolname{} using 21 open-source libraries from the \utopia{} benchmark and 21 libraries from the \promptfuzz{} benchmark suite (with two libraries overlapping between the two suites), addressing four key research questions (RQs).
First, we assessed the effectiveness of \toolname{} in fuzz driver generation (RQ1). Second, we compared \toolname{} with several baseline techniques, including manually written drivers in \ossfuzz{}~\cite{ossfuzz}, the unit-test-based synthesis tool \utopia{}\cite{jeong2023utopia}, and two recent LLM-based methods, \promptfuzz{}\cite{lyu2023prompt} and \ckgfuzzer{}~\cite{xu2024code} (RQ2). Third, we analyzed the contributions of each component in \toolname{} to its overall performance (RQ3). Finally, we conducted a case study to examine the bugs detected and the corresponding fuzz drivers generated by \toolname{} (RQ4).
Experimental results show that \toolname{} consistently outperforms all baselines. Specifically, \toolname{} achieves higher region coverage than \utopia{} on 16 out of 21 libraries. It also improves overall branch coverage by 1.62$\times$, 1.50$\times$, and 1.89$\times$ compared to \ckgfuzzer{}, \promptfuzz{}, and \ossfuzz{}, respectively. Furthermore, \toolname{} discovered 33 previously unknown bugs in widely used libraries, 3 of which have been assigned CVEs.

The main contributions of this paper are as follows:
\begin{itemize}
    \item \textbf{API Combination Constraints.} We devise explicit and implicit constraints to formalize the validity and rationality of API combinations. Furthermore, we develop an effective approach to extract and solve these constraints.
    \item \textbf{Dual Scheduling Framework.} To the best of our knowledge, we are the first to propose the perspective of treating driver generation and execution as an online optimization problem and dynamically scheduling them.
    \item \textbf{\toolname{}.} We implement a prototype of this approach and conduct rigorous experiments. The experimental results demonstrate that \toolname{} outperforms all state-of-the-art baseline tools (\ossfuzz{}, \utopia{}, \promptfuzz{} and \ckgfuzzer{}), each based on different driver synthesis techniques.

\end{itemize}

%% file: figures/valid_but_irrational.tex
\begin{figure}[t]
    \centering

    \includegraphics[width=0.45\textwidth]{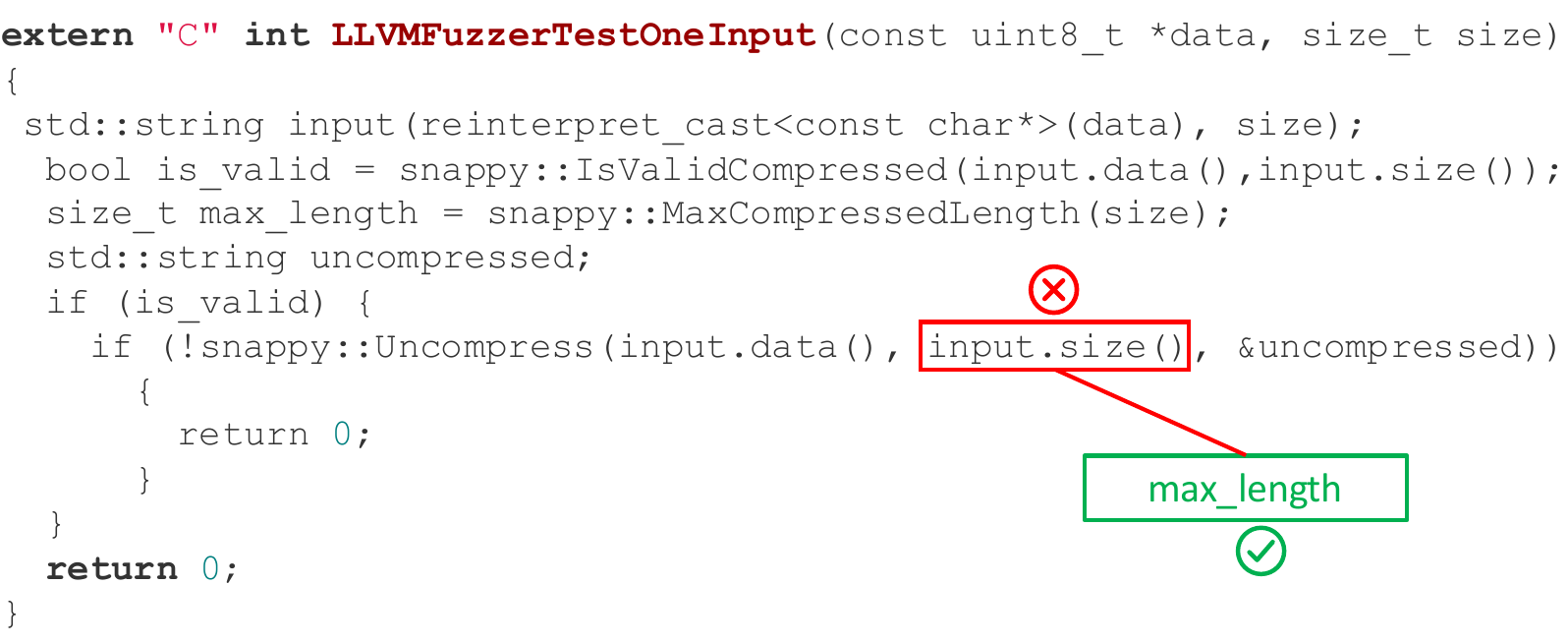}
    \caption{A valid fuzz driver but irrational for \textit{snappy}}
    \label{fig:valid but irrational}
\end{figure}

%% file: figures/overview.tex
\begin{figure*}[t]
    \includegraphics[width=0.99\textwidth]{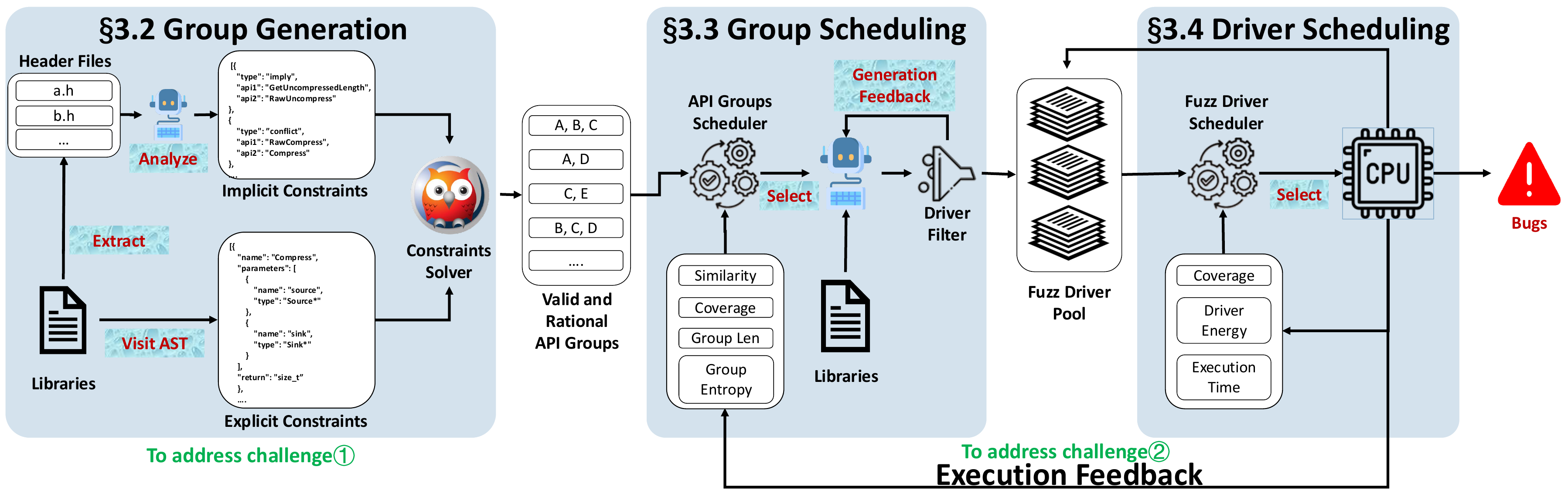}
    \caption{Overview}
    \label{fig:overview}
\end{figure*}

%% file: sections/1_background.tex
\section{Preliminaries}

This section provides the background of this paper, covering the API combination issues and problem definitions.




\subsection{API Combination}

\boldparagraph{Validity and Rationality of Combination.}
As shown in \autoref{fig:valid but irrational}, the driver calls three APIs: \textit{IsValidCompressed}, \textit{MaxCompressedLength}, and \textit{Uncompress} sequentially.
Specifically, there are data dependencies and control flow dependencies between them. 
These dependencies require type matching; otherwise, they result in compilation errors in fuzz drivers.
Therefore, a valid API combination should ensure type consistency to allow successful compilation. 
However, not all valid combinations are rational. 
Although these three APIs can be combined to form a valid fuzz driver, it is a completely incorrect usage of \textit{Snappy}. Since the compressed data from \textit{IsValidCompressed} and the compressed length from \textit{MaxCompressedLength} are unaligned, an runtime error occurs in \textit{Uncompress} when this fuzz driver is executed. 
Unfortunately, the improper usage of APIs causes this error that cannot be identified a bug by library maintainers. 
Existing work~\cite{babic2019fudge, ispoglou2020fuzzgen, lyu2023prompt, jeong2023utopia, zhang2021apicraft} concentrates on synthesizing valid and efficient drivers but lacks a focus on rationality.

To make things worse, a library exports numerous APIs, and many of them can be combined. 
As a result, the number of combinations will grow exponentially if we aim to explore all possible combinations.
For example, there are 15 APIs in the library \textsc{snappy}, and the number of possible combinations is 32,768, making it impractical to generate drivers for each combination. 
Therefore, to improve the fuzzing performance, we need to efficiently schedule combinations and drivers.

\subsection{Problem Definition.}
Automatic library fuzzing can be defined as an online optimization problem aiming to maximize the execution time of rational API combinations. 
Let us begin with an arbitrary target library denoted as $L$, which consists of a set of APIs denoted as $A$. A set $D$ of all fuzzing drivers, regardless of whether they are correct or wrong from $L$. 

\begin{definition}[Validity]
We say that API $P$ depends on $Q$ when the output of $Q$ is the input of $P$, denoted as $deps(P, Q)$.
An API combination $C$ consists of multiple APIs,
we say its a valid API combination if $C$ has the property:
$$\forall f \in C.\ \exists g \in C.\ deps(f,g)\ ||\ deps(g,f),\ where\ f \neq g$$
\end{definition}

\begin{definition}[API Implication and Conflict]
An API $P$ is said to imply API $Q$ if the library's usage convention mandates that $Q$ must be called after $P$, represented as $P \leadsto Q$. Conversely, $P$ is said to conflict with $Q$ if both APIs cannot be invoked together within a single unit of functionality, denoted as $P \otimes Q$.
\end{definition}

\begin{definition}[Rationality]
A valid API combination $C$ is said to be rational within a library's usage convention, given implication constraints $I$ and conflict constraints $O$, if the following conditions hold:
\begin{align*} 
\forall i \in I. \exists q. (\exists p \in C. (p \leadsto q) = i)   &\rightarrow (\exists g \in C. g = q) \\
\forall o \in O. \exists q. (\exists p \in C. (p \otimes q) = o)    &\rightarrow (\forall g \in C. g \neq q)
\end{align*}
The corresponding fuzz driver is referred to as a rational driver and is denoted by $D_R$. Bugs identified using $D_R$ are more likely to be accepted by library maintainers.
\end{definition}

At each driver scheduling point $i$, the fuzzer selects $N$ fuzzing drivers $\sum_{j=1}^{N}D_R[u_i(j)]$ to execute from the fuzzing driver pool, where $u_i$ represents the index of the chosen drivers. Based on the scheduling process, the objective is to maximize the coverage $cov(*)$ in $L$ by rational fuzzing drivers in $D$ across $K$ stages in one fuzzing campaign. This can be formally represented as follows:

\begin{equation}
  \begin{aligned}
    Maximize \sum_{i=1}^{K}\sum_{j=1}^{N}[cov(D_R({u_i[j]}))], \\
  \end{aligned}
\end{equation}

We frame this challenge as an online optimization problem~\cite{DBLP:journals/corr/abs-1909-05207}, as the fuzzer can only observe the $cov(*)$ output of API combinations after applying LLMs and executing fuzz drivers. In the context of online optimization for library fuzzing, decisions must be made at each stage without full knowledge of the objective function’s values for subsequent stages.
The primary goal of our fuzzer is to maximize cumulative coverage of rational fuzz drivers over a sequence of stages. This results in an online optimization problem, where identifying the optimal sequence of API combinations and fuzz drivers is crucial to achieving the highest coverage throughout the fuzzing campaign.

%% file: sections/2_0_design.tex
\section{Proposed Approach}
\input{figures/visitor}
\input{figures/prolog}
\input{figures/implicit_constraints_prompt}
\input{figures/driver_generation_prompt}
\input{figures/repair_driver_prompt}
In this section, we begin by introducing the workflow of \toolname{}.
Consequently, we detail how \toolname{} generates API Group from a library by an LLM and the Prolog solver. After that, we show the \textsc{Group Scheduler} (\groupschedule{}) and \textsc{Driver Scheduler} (\driverschedule{}) that address the challenge of API combinatorial explosion.

\input{sections/2_1_overview}
\input{sections/2_2_group_generation}

\input{sections/2_3_group_schedule}

\input{sections/2_4_driver_schedule}

%% file: figures/visitor.tex
\begin{algorithm}[t]
\caption{The Explicit Constraint Extraction Algorithm}\label{alg:The Explicit Constraint Extraction Algorithm}

\begin{algorithmic}[1]
\Require $\mathcal{L}$: Static Library path, $\mathcal{D}$: Source files path
\Ensure $\mathcal{A}$: Set of type-based constraints, $\mathcal{S}$: Set of valid symbols

\Function{Analyze}{$\mathcal{L}, \mathcal{D}$}
    \State $\mathcal{M} \gets \text{RunCommand}(\text{nm}, \mathcal{L})$
    \State $\mathcal{D}_{\text{sym}} \gets \text{Demangle}(\mathcal{M})$
    \State $\mathcal{S} \gets \text{RemoveModifiers}(\mathcal{D}_{\text{sym}})$
    \State $\mathcal{F} \gets \{f \mid f \in \mathcal{D}, \text{ext}(f) \in \{\text{h}, \text{hpp}, \text{c}, \text{cpp}, \text{cc}\}\}$
    \State $\mathcal{S}^* \gets \{s \in \mathcal{S} \mid \exists f \in \mathcal{F}, s \in \text{Content}(f)\}$ 
    
    \State $\mathcal{A} \gets \emptyset$
    \ForAll{$f \in \mathcal{D}$, $\text{ext}(f) \in \{\text{c}, \text{cpp}, \text{cc}\}$}
        \State $\text{AST} \gets \text{GenerateAST}(f)$
        \ForAll{nodes $n \in \text{AST}$}
            \If{$n.\text{Kind} \in \{\text{Function}, \text{Method}, \text{Constructor}\}$}
                \State $t_r \gets n.\text{ResultType}$
                \State $\mathcal{P} \gets [(p.\text{Type}, p.\text{Name}) \mid p \in n.\text{Parameters}]$
                \State $\mathcal{A} \gets \mathcal{A} \cup \{ \{ n.\text{Name}, n.\text{Signature}, t_r,\mathcal{P} \} \}$
            \EndIf
        \EndFor
    \EndFor
    
    \State \text{Save} $\mathcal{S}^*$ \text{and} $\mathcal{A}$ \text{to JSON}
    \State \Return $\mathcal{A}, \mathcal{S}^*$
\EndFunction

\end{algorithmic}

\end{algorithm}

%% file: figures/prolog.tex
\begin{figure}[t]
\scriptsize
\centering

\begin{subfigure}{0.48\textwidth}
        
        \begin{tabular}{p{0.41\textwidth}p{0.5\textwidth}}
        \toprule
        \textbf{Constraints}  & \textbf{Description}  \\ 
        \midrule
        parameter\_types($f \in F$, $ts \in [T]$). & $ts$ is the parameter types of $f$ \\
        return\_type($f \in F$, $t \in T$). & $t$ is the return type of $f$ \\
        \midrule
        imply($f_1 \in F$, $f_2 \in F$). & A group must contain $f_2$ if $f_1$ in it                            \\
        conflict($f_1 \in F$, $f_2 \in F$).  & A group can not contain both $f_1$ and $f_2$                     \\
        \bottomrule
        \end{tabular}
    \caption{Definitions of constraints.}
    \label{fig:constraints}
\end{subfigure}
\vspace{0.2cm}

\begin{subfigure}{0.5\textwidth}
        \begin{tabular}{p{0.92\textwidth}}
        \toprule
        \textbf{Prolog rules ($\mathcal{M}$ indicates the built-in predicate member of prolog)}    \\ 
        \midrule
        same\_parameter($t \in T$, $g' \in G$) :- $\mathcal{M}$($f$, $g'$), parameter\_types($f$, $ts$), $\mathcal{M}$($t$, $ts$).                            \\
        same\_return($t \in T$, $g' \in G$) :- $\mathcal{M}$($f$, $g'$), return\_type($f$, $t$). \\
        \midrule
        depends($f \in F$, $g' \in G$) :- parameter\_types($f$, $ts$), $\mathcal{M}(t, ts)$, same\_return($t$, $g'$).   \\
        depends($f \in F$, $g' \in G$) :- parameter\_types($f$, $ts$), $\mathcal{M}(t, ts)$, same\_parameter($t$, $g'$).   \\
        depends($f \in F$, $g' \in G$) :- return\_type($f$, $t$), same\_parameter($t$, $g'$).  \\
        \midrule
        solve\_explicit($[]$, $g \in G$).\\
        solve\_explicit($[f  | fs ] \in G$, $g \in G$) :- delete($g$, $f$, $g'$), depends($f$, $g'$), solve\_explicit($fs$, $g$).\\
        \midrule
        sat\_explicit($g \in G$) :- solve\_explicit($g$, $g$). \\
        \bottomrule
        \end{tabular}
    \caption{Prolog rule definitions of the explicit constraints solver.}
    \label{fig:Prolog rule definitions of explicit}
\end{subfigure}
\vspace{0.2cm}

\begin{subfigure}{0.5\textwidth}
        \begin{tabular}{p{0.93\textwidth}}
        \toprule
        \textbf{Prolog rules ($\mathcal{M}$ indicates the built-in predicate member of prolog)}    \\ 
        \midrule
        solve\_implicit($g \in G$, $(f_1, f_2) \in$ imply) :- $\mathcal{M}$($f_1$, g) -> $\mathcal{M}$($f_2$, g). \\ 
        solve\_implicit($g \in G$, $(f_1, f_2) \in$ conflict) :- not($\mathcal{M}$($f_1$, g), $\mathcal{M}$($f_2$, g)). \\
        \midrule
        sat\_implicits($g \in G$, $[]$). \\
        sat\_implicits($g \in G$, $ [i | is ] \in [I]$) :- solve\_implicit($g$, $i$), sat\_implicits($g$, $is$).                             \\
        \bottomrule
        \end{tabular}
    \caption{Prolog rule definitions of the implicit constraints solver.}
    \label{fig:Prolog rule definitions of implicit}
\end{subfigure}

    \caption{The Prolog solver setup}
    \label{fig:Prolog setup}
\end{figure}

%% file: figures/implicit_constraints_prompt.tex
\begin{figure}[t]
    \centering

    \includegraphics[width=0.45\textwidth]{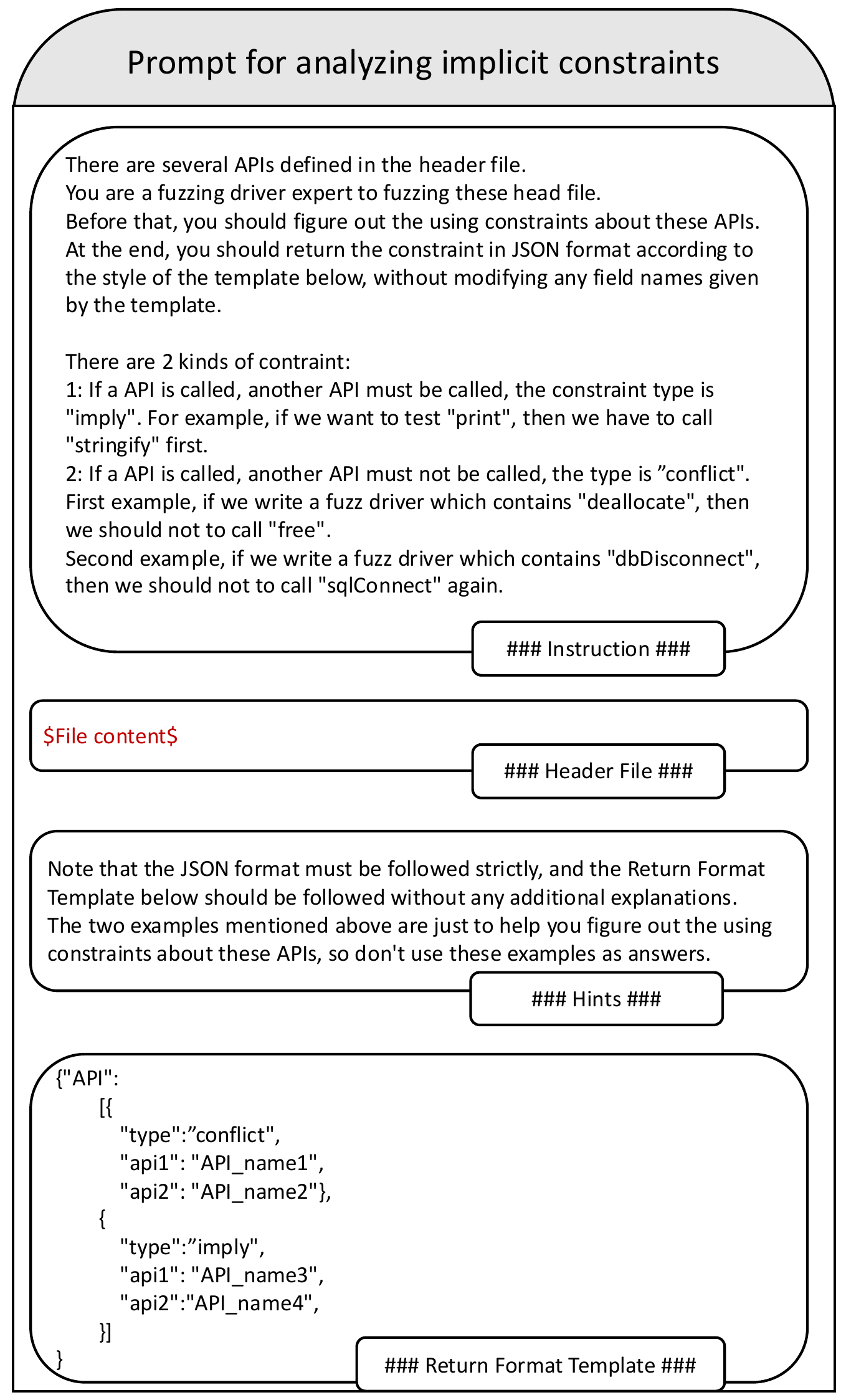}
    
    \caption{The prompt of analyzing implicit constraints.}
    \label{fig:The prompt of analyzing implicit constraints}
\end{figure}

%% file: figures/driver_generation_prompt.tex
\begin{figure}[t]
    \centering
    \includegraphics[width=0.43\textwidth]{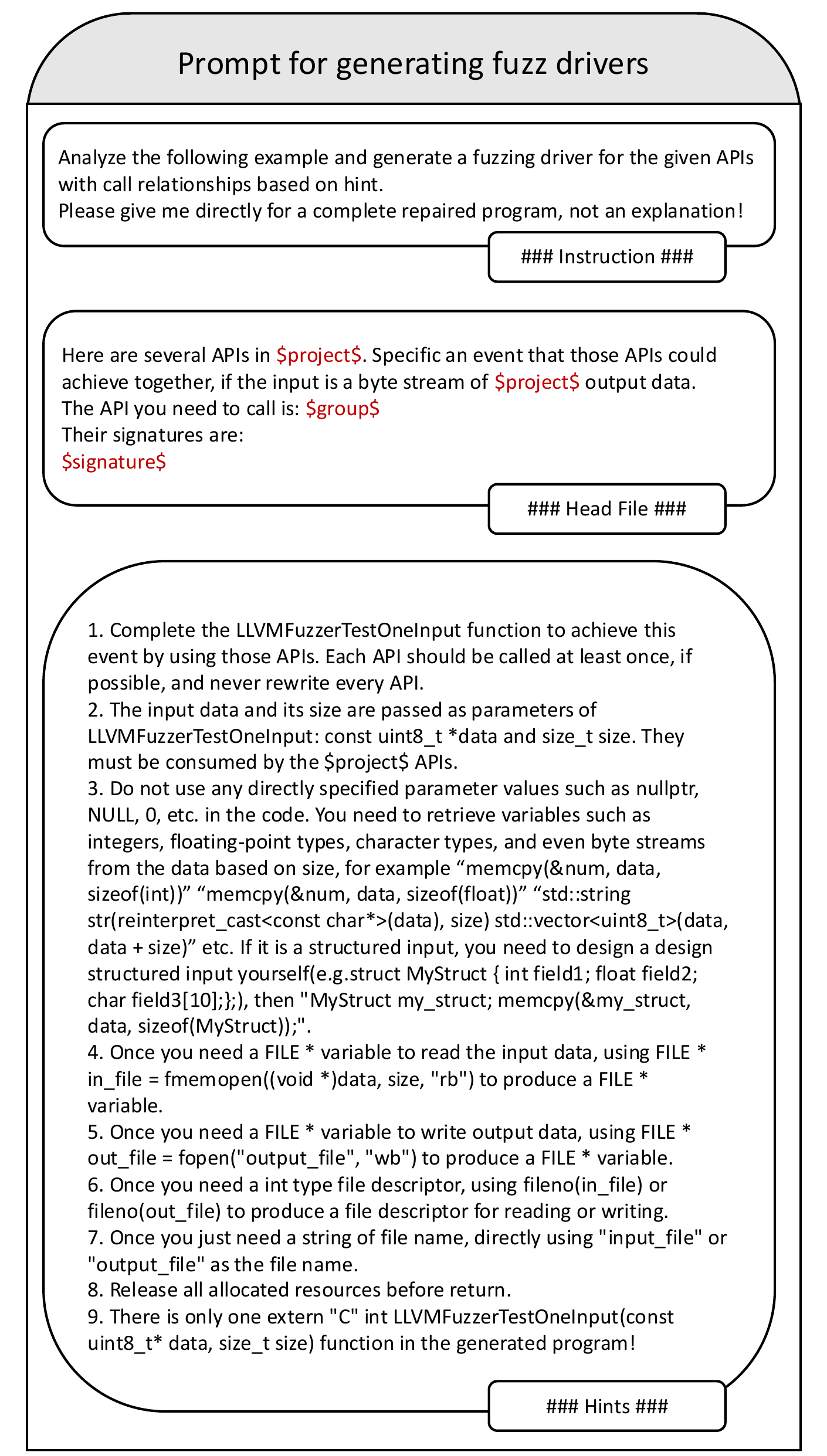}
    \caption{The prompt of generating fuzz drivers.}
    \vspace{-10pt}
    \label{fig:The prompt of generating fuzz drivers}
\end{figure}

%% file: figures/repair_driver_prompt.tex
\begin{figure}[t]
    \centering

    \includegraphics[width=0.45\textwidth]{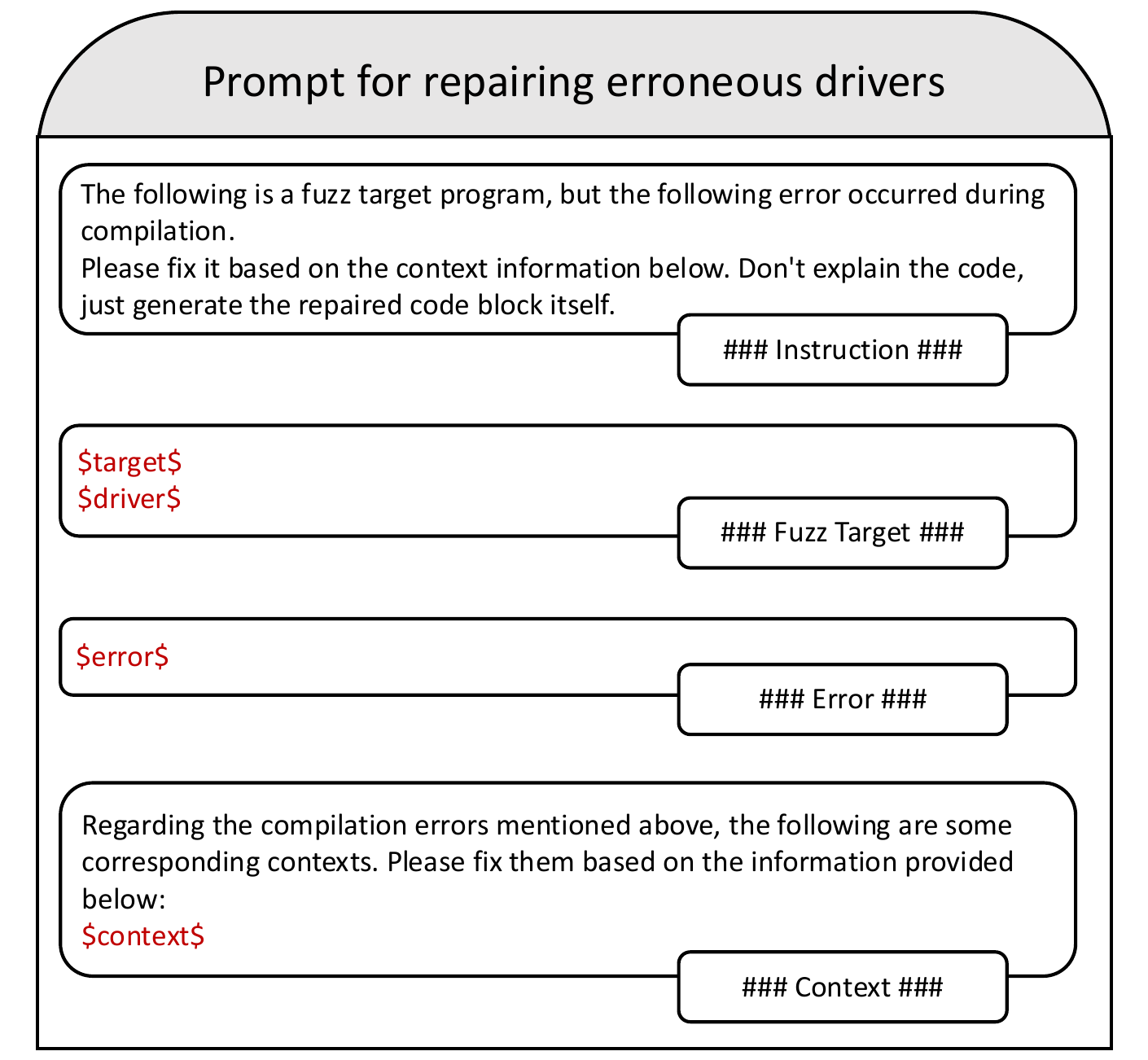}
    
    \caption{The prompt of re-generating fuzz drivers.}
    \label{fig:The prompt of re-generating fuzz drivers}
\end{figure}

%% file: sections/2_1_overview.tex
\subsection{Overview}

\autoref{fig:overview} details the workflow of \toolname{}.
Overall, it receives the entire project files of a fuzzing subject library, and generate numerous fuzz drivers to execute them in a loop.
\toolname{} consists of three components: 
{\large \textcircled{\small 1}}\textit{Group Generation}, {\large \textcircled{\small 2}}\textit{Group Scheduling} and {\large \textcircled{\small 3}}\textit{Driver Scheduling}.

In component{\large \textcircled{\small 1}}, \toolname{} extracts all the header files to obtain the implicit constraints by an LLM and performs AST visitors to collect all explicit constraints. Then, the two constraints are sent to a constraints solver written in Prolog. After the solver finishes, this component outputs numerous rational API groups for the next scheduler.  
In component {\large \textcircled{\small 2}}, it accepts all rational API groups as a driver generation corpus for the LLM. To select which groups can be used to generate the corresponding driver, the component \groupschedule{} computes the score based on four metrics. Consequently, \toolname{} constructs a generation prompt with library and group information for the LLM. \groupschedule{} then queries the LLM to generate fuzz drivers in a loop until reaching the loop times limitation. Filtered fuzz drivers are saved in the fuzz driver pool for the next component. Meanwhile, the component \driverschedule{} selects some fuzz drivers from the pool based on three metrics, and these drivers begin fuzzing for a CPU time slicing. To help optimize the scheduling, the execution results are sent as feedback information for \groupschedule{} and \driverschedule{}.

%% file: sections/2_2_group_generation.tex
\subsection{API Group Generation}
In \toolname{}, two types of constraints are proposed to formalize the valid and rational API groups.
Furthermore, we develop an automatic constraints collection approach and an effective solver for them.
To describe the constraints, we encode them into a Prolog representation. 
Prolog is a logical programming language, allowing us to conveniently express constraints and solve them. 
We use the following domains to encode the information about the constraints as Prolog facts, 
noting $F$ as the set of API functions and $T$ as the set of formal types. 
We show an overview of the constraints that we use to check for valid and rational API groups in \autoref{fig:constraints}. 
Informally, explicit constraints refer to the API types, while implicit constraints describe the relationships between two APIs, such as implication and conflict.

\boldparagraph{Explicit Constraints Collection.}
\toolname{} utilizes an AST visitor to extract all the function formal types in header files and generates explicit constraints. As shown in \autoref{alg:The Explicit Constraint Extraction Algorithm}, this process starts by extracting mangled symbols from the static library using the `nm` command, followed by demangling and simplification to generate a symbol set $\mathcal{S}$ (lines 2–3). These symbols are then filtered to identify the valid subset $\mathcal{S}^*$ by matching them against the contents of relevant source files $\mathcal{F}$, which include header and implementation files (lines 4–6). Next, the source files in the directory are parsed to generate Abstract Syntax Trees (ASTs), and nodes representing functions, methods, or constructors are traversed (lines 7–19). For each node, a matching process aligns its attributes, such as the name and namespace, with symbols in $\mathcal{S}$ to extract detailed API information. This includes function signatures, parameter types, and return types, covering not only primitive C/C++ types but also user-defined types (lines 10–16). Finally, both the filtered symbol set $\mathcal{S}^*$ and the extracted API details $\mathcal{A}$ are saved in JSON format. 
For example, the explicit constraints of the function \inlinecpp{snappy::Compress} in Snappy include its signature: \inlinecpp{size_t Compress(snappy::Source* reader, snappy::Sink* writer)}, its return type \inlinecpp{size_t}, and its parameters: \inlinecpp{reader} of type \inlinecpp{snappy::Source}, and \inlinecpp{writer} of type \inlinecpp{snappy::Sink}. 
Using explicit constraints, we can easily check whether the APIs in a group can be combined together as a type-valid calling sequence.


\boldparagraph{Implicit Constraints Collection.}
Moreover, to obtain the implicit constraints from code, \toolname{} leverages an LLM to analyze the library under test.
As shown in \autoref{fig:The prompt of analyzing implicit constraints}, 
the prompt template requires the C/C++ header files to fill in the \textit{Header file} section. 
To formalize the relationship between any APIs, the prompt requires LLMs to analyze and summarize two kinds of implicit constraints for each library: \inlinecpp{imply(foo, bar)} and \inlinecpp{conflict(foo, bar)}. The former indicates that \inlinecpp{bar} must be included in an API group if \inlinecpp{foo} exists. The latter means that \inlinecpp{foo} and \inlinecpp{bar} cannot exist in the same API group. 
These two implicit constraints describe the basic relationship between two APIs with capability of composition.

\boldparagraph{Constraints Solver.} 
Generally, the solver generates all API permutations and filters them based on the collected constraints.
The filter details of the solver are shown in \autoref{fig:Prolog rule definitions of explicit}, the predicate \inlinecpp{sat\_explicit} takes an API group and calls \inlinecpp{solve\_explicit} to solve it with the same two groups: one indicates unchecked list, and the other is the total list. 
Consequently, \inlinecpp{solve_explicit} checks whether each function depends on others by utilizing the predicate \inlinecpp{depends}, which ensures that a function's input and output types are present in the remaining functions within the group $g$.
Notably, different conventions are often employed depending on the API or individual functions, where parameters may serve as inputs, outputs, or both. To address this, \inlinecpp{depends} allows parameter types to match either the parameter types or the return types of other functions.
If a function's input and output types are unrelated to others, the function is identified as an orphan in this group, making the group invalid and subject to removal. 
Otherwise, \inlinecpp{solve\_explicit} continues to check the remaining functions until all the functions in this group are checked.
In \autoref{fig:Prolog rule definitions of implicit}, \toolname{} identifies a group as rational once it satisfies all the implicit constraints. 
More specifically, we use the predicate \inlinecpp{->/2}~\cite{->} to model the \inlinecpp{imply} constraint. The rule $\mathcal{M}$($f_1$, $g$) $\to$ $\mathcal{M}$($f_2$, $g$) indicates that if API $f_1$ exists in the API group $g$, then API $f_2$ must also be in the same group.
Similarly, we model the \inlinecpp{conflict} constraint using the predicate \inlinecpp{not/1}~\cite{not}, which denotes that the API group $g$ violates the implicit constraints if both APIs $f_1$ and $f_2$ exist in $g$.
After execution of the solver, it generates all the valid and rational API groups for the next component.

%% file: sections/2_3_group_schedule.tex
\subsection{Group Scheduling}
\boldparagraph{Group Priority.}
Since the API groups are too large to generate all the corresponding fuzz drivers, \toolname{} tries to select the optimal ones in each loop step.
The Group Scheduler (\groupschedule{}) accepts the entire valid and rational API group list, and utilizes multi-objective optimization to prioritize these groups for higher code coverage and fuzzing efficiency. To construct the objectives, we devise four metrics: \textit{similarity}, \textit{code coverage}, \textit{group length}, and \textit{group entropy}. The reasons for selecting these metrics are as follows:
\begin{itemize}
    \item \textit{Similarity}. \groupschedule{} tends to select similar groups when the previous group achieves high code coverage. Inspired by the intuition behind coverage-guided seed selection, similar API groups are treated as mutations of the one achieving high coverage. 
    Given a set of previously selected API groups $\mathbb{S}$ and the corresponding code coverage $C_{s \in \mathbb{S}}$, the similarity score of a candidate group $g$ can be calculated using \textsc{Jaccard Similarity Coefficient}~\cite{khoshmanesh2018role} as:
        $$ \sum_{s \in \mathbb{S}} \text{Jaccard}(s, g) \times C_s $$
    \item \textit{Code coverage}. Groups with higher coverage are more likely to explore new code areas and detect additional bugs. Generating drivers from these groups more frequently can improve fuzzing performance.
    \item \textit{Group length}. 
    Since few unit functionalities require more than five APIs~\cite{jeong2023utopia, lyu2023prompt}, and a shorter group length typically improves the success rate of LLM generation, we limit the group length to five, following prior work.
    \item \textit{Groups entropy}. To balance the number of all APIs, we calculate the entropy of the groups and select those leading to higher entropy. Given a set of API groups $\mathbb{D}$ in the fuzz driver pool, a set of exposed APIs $\mathbb{F}$, and the frequency of occurrence $p_f$ for each API $f$, the group entropy is computed as follows:
    $$  
    \text{Entropy}(\mathbb{D}) = -\sum_{f \in \mathbb{F}} p_f \log_2(p_f)  
    $$
\end{itemize}

Unfortunately, these metrics are often in conflict with each other. For example, as \textit{similarity} increases, \textit{Groups entropy} tends to decrease. This dynamic may inadvertently limit the exploration of new or untested code regions, as it emphasizes familiar pathways over diversification.
To address this issue, \toolname{} aims to minimize \textit{Group length} while maximizing the other metrics, employing a multi-objective optimization approach to achieve the best testing performance.
For efficient optimization, we adopt the nondominated sorting algorithm from \textsc{CEREBRO}~\cite{li2019cerebro}. The core idea is to identify the most critical edge API groups (Pareto frontier) by ranking the groups and iteratively selecting them from lower to higher ranks.


\boldparagraph{Fuzz Driver Generation.}
Initially, \toolname{} randomly selects an API group to bootstrap the fuzzing process. Subsequently, a batch of API groups is selected to generate drivers. Once the API groups are selected, they are used to instantiate the prompt template and then sent to an LLM to synthesize fuzz drivers. As shown in \autoref{fig:The prompt of generating fuzz drivers}, the prompt template consists of three parts: The first part (Instruction) describes the primary task; The second part (Given API) replaces the placeholders \textit{project}, \textit{group}, and \textit{signature} with the project name, selected APIs, and its signature, respectively. In addition, we summarize some common generation errors of LLMs and include nine prompt hints in the third part (Hint).

However, due to LLM's hallucination~\cite{zhang2023hallucination}, the generated fuzz drivers might not satisfy the given instructions and hints. 
Therefore, \groupschedule{} sets up a driver filter to assess whether the driver meets the requirements statically and dynamically. First, the filter checks whether the driver contains the APIs required by the API group. Then, the driver is executed, and \groupschedule{} rejects those that cannot be compiled or crash within fifteen seconds (short-term fuzzing). In previous work~\cite{10.1145/3650212.3680355}, the short-term fuzzing threshold was set to one minute, achieving zero false positives. In this study, we reduce the threshold to fifteen seconds to further minimize the likelihood of false positives.
Consequently, \groupschedule{} sends them back for regeneration with error feedback. Furthermore, to eliminate the compilation errors, we design an information retriever to search for supplementary information. After receiving the error information, the retriever first matches the focus information using regular expressions, as detailed in \autoref{sec: Regular Expression of Retriever}. This process includes handling cases such as unmatched function and member names, undeclared identifiers, and incompatible type names. Afterwards, the definition of the corresponding focus information, such as function bodies, constant values, etc., will be retrieved through string matching in the source code. Finally, combine the retrieved information as the context for error feedback prompt.
After that, \groupschedule{} places it in our repair prompt, as shown in ~\autoref{fig:The prompt of re-generating fuzz drivers}, with type information (\textit{target}) and previous driver information (\textit{driver}).
Once some drivers are accepted by the filter, they are added to the fuzz driver pool and await driver scheduling.

%% file: sections/2_4_driver_schedule.tex
\subsection{Driver Scheduling}
To efficiently use computational resources, we devise the \textit{Driver Scheduling} (\driverschedule{}) component to suspend and resume fuzz drivers in each loop step.



We propose three metrics to compute a saturation score to measure whether the coverage of a driver is saturated. The first metric is \textit{driver energy}.
Once a driver suspends after a CPU time slice, its energy is reduced accordingly. \driverschedule{} tends to select drivers based on their remaining energy. 
The other two metrics are \textit{code coverage} and \textit{accumulated execution time}. We calculate the $score = \frac{code\ coverage}{execution\ time}$ for each driver. 
Consequently, \driverschedule{} leverages the \textit{Roulette Wheel Selection} algorithm ~\cite{lipowski2012roulette} on the score to select a batch of drivers. 
Thus, a higher score indicates a greater opportunity to be chosen. 
Notably, drivers with zero execution time will be prioritized since their score is infinite. Conversely, if the coverage of a driver fails to increase for a long time, its score will continually decrease, reducing its chances of being chosen.




After each CPU time slice, \driverschedule{} selects some drivers based on the aforementioned metrics and begins to execute or resume them. During execution, \toolname{} collects coverage and updates the execution feedback after each CPU time slice. 
To avoid duplicate bug detection, drivers are removed from the fuzz driver pool once a bug is found; otherwise, they remain in the pool and await scheduling.

%% file: sections/5_0_eval.tex
\section{Evaluation}
\boldparagraph{Research Questions.} We conduct experiments to evaluate \toolname{} by tackling the following questions:
\begin{itemize}
    \item \textbf{Effectiveness:} How effective is the driver generation capability of \toolname{}?
    \item \textbf{Advancement:} Is \toolname{} better than the state-of-the-art baseline fuzz driver generation techniques? 
    \item \textbf{Necessity:} What are the contributions of each component of \toolname{}?
    \item \textbf{Case Study:} What are the detected bugs and corresponding fuzz drivers?
\end{itemize}

\boldparagraph{Implementation.}
We implemented \toolname{} with 7K lines of code (excluding comments or blank lines), among which 5.5K are python code to analyze the libraries and schedule the generation and running of the fuzzing's driver, whereas the remaining 1.5K Loc are prolog and shell sripts to solve group combinations and simplify the analysis and compilation process of the entire library respectively. The code for API information extraction, compilation filtering and context retrieval leverages the analysis framework of LLVM/Clang and the corresponding version of libclang's Python package.

\boldparagraph{Evaluation Setup.}
Our experiments are performed on a Linux server with Intel(R) Xeon(R) Gold 6240 CPU @ 2.60GHz and distribution Ubuntu 20.04. \toolname{} leverage \textit{GPT-3.5-turbo-0125} API of OpenAI as the LLM, and set the temperature parameter to 1. In addition, \toolname{} repeatedly queries LLMs to generate drivers when they are filtered due to missing required APIs or compilation failures. To balance the trade-off between failure rate and token cost, we set the maximum number of repeated queries to
four. Experimental results on the overall failure rate with different query times settings are presented in \autoref{sec: Failure Analysis Patterns of Driver Generation}.

To demonstrate the performance of \toolname{}, we compared the code coverage of libraries against the human-centric fuzz drivers in \ossfuzz{}, the unit-test-based fuzz driver generator \utopia{} and the state-of-the-art LLM-based fuzzing tool \promptfuzz{} and \ckgfuzzer{} .
Notably, we evaluate \toolname{} under stricter resource constraints: while \utopia{} were run for 100 per-core hours on 112-core CPUs, we limit \toolname{} to 20 per-core hours on 5 cores. For \promptfuzz{} and \ckgfuzzer{}, substantial pre-fuzzing time was spent generating drivers, each of which was executed for 24 hours on separate CPU cores. To ensure fairness given architectural differences, we configure \toolname{}'s fuzzing campaign to 16 cores × 24 hours per library.

\boldparagraph{Benchmarks.}
Since \utopia{} is designed to work with libraries that leverage the \textsc{GoogleTest}~\cite{gtest} framework, while \promptfuzz{} is a general library fuzzing tool, we cannot evaluate them under a single unified configuration. Therefore, to compare with \utopia{} and \promptfuzz{}, we use all the 21 libraries from the original \utopia{} paper and all the 12 libraries from the original \promptfuzz{} paper.

To compare against the baseline techniques using the same standard, we use region coverage for \utopia{} and branch coverage for \promptfuzz{} and \ckgfuzzer{}. Region coverage is natively supported in LLVM, whereas branch coverage for \promptfuzz{} and \ckgfuzzer{} is custom-implemented. Accordingly, we evaluate \toolname{} using both region and branch coverage metrics for completeness.

\boldparagraph{Metrics.} We utilize three important metrics to evaluate \toolname{} as follows:
\begin{itemize}
    \item \textbf{Stillborn rate.} To measure the valid driver generation capability of \toolname{}, we introduce the stillborn rate (SR) for \groupschedule{}. 
    Given the current size of the compilable drivers ($|CD|$) and the number of LLM queries ($\#Queries$), the SR can be calculated as follows:
    \begin{equation}
        SR = 1 - \frac{|CD|}{\#Queries} \times 100\% 
    \end{equation}
    In our experiments, stillborn artifacts are drivers with compilation errors or experience early crashes. 
    Thus, the stronger the generation capability of \toolname{}, the lower this rate will be.

    \item \textbf{Region coverage and Branch Coverage.} 
    To compare against the baseline techniques using the same standard, we use region coverage for \utopia{} and branch coverage for \promptfuzz{}. 
    Region coverage is a built-in feature in LLVM, while branch coverage in \promptfuzz{} is custom-implemented. 
    Therefore, we evaluate \toolname{} using both the built-in region coverage and the custom branch coverage from \promptfuzz{}.
\end{itemize}

\input{figures/coverage}
\input{figures/driver_generation}
\input{figures/comparison_with_utopia}

\input{figures/comparison_with_promptfuzz}
\input{figures/ablation}

\input{sections/5_1_rq1}

\input{sections/5_2_rq2}
\input{figures/ablation_coverage}
\input{sections/5_3_rq3}
\input{sections/5_4_rq4}

\input{figures/bug1}
\input{figures/bug2}

%% file: figures/coverage.tex
\begin{figure*}[t]

    \begin{subfigure}{\textwidth}
            \includegraphics[width=1\textwidth]{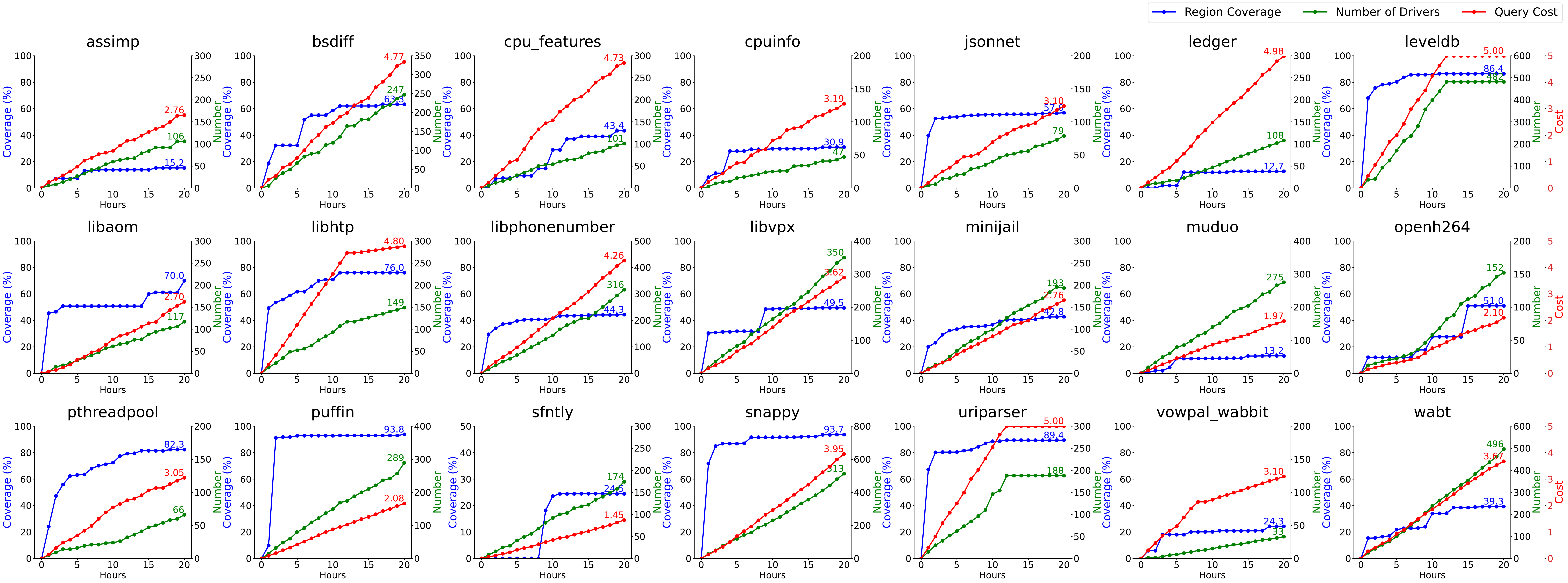}
            \caption{The coverage on the libraries from \utopia{}.}
            \label{fig:ut}
    \end{subfigure}

    \begin{subfigure}{\textwidth}
            \includegraphics[width=1\textwidth]{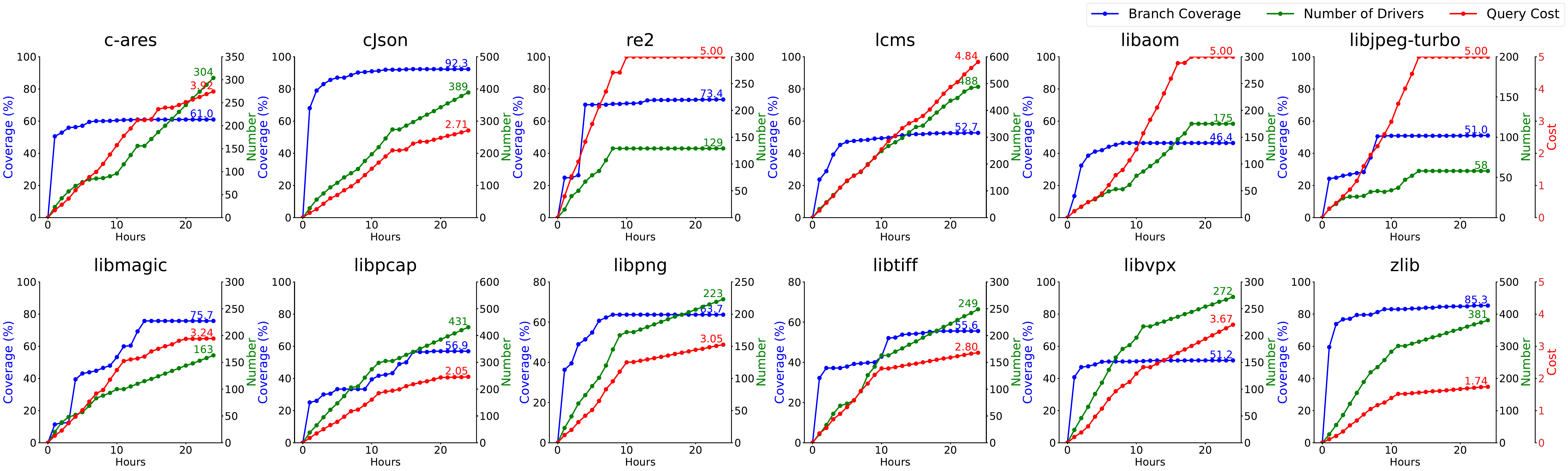}
            \caption{The coverage on the libraries from \promptfuzz{}.}
            \label{fig:pf}
    \end{subfigure}

    \caption{Regions and branches covered by \toolname{} with number of generated drivers and query cost.}
    \label{fig:coverage}

\end{figure*}

%% file: figures/driver_generation.tex
    \begin{table}[t]
        \centering
        \caption{Group and driver generation.}
        \resizebox{.48\textwidth}{!}{
             \begin{tabular}{c|l|ll|l|l|l|l}
                \toprule
                \multicolumn{2}{c|}{\textbf{Library}}  & \multicolumn{2}{c|}{\textbf{\#Constraints}} & \multirow{2}{*}{\textbf{\#Groups}} & \multirow{2}{*}{\textbf{SR}} & \multirow{2}{*}{\textbf{\#Drivers}} & \multirow{2}{*}{\textbf{Cost}} \\
                \cmidrule{1-4}
                \textbf{Category}              &     \textbf{Name}               &   \textbf{I}                              &  \textbf{E}                   &          &                      & \\
                
                \midrule
                \multirowcell{10}{Multiple \\ Media}  &assimp             & 9 & 2484      & 170833 & 0.86  &106  &2.76 \\
                                    &libaom              & 29&47       & 27485 &0.86   &175 &5.00 \\
                                    &libvpx              & 8&40         & 116846 &0.84   &272 &3.66 \\
                                    &openh264            & 5&1123        & 453012 &0.75  &152  &2.10 \\
                                    &sfntly              & 24&1103     & 12541 &0.89  &174  &1.45 \\
                                    &vowpal\_wabbit      & 12&1183 & 36262 &0.94  &33   &3.10\\
                                    &lcms          & 3&291 & 68478 & 0.73 &488  &4.83\\
                                    &libjpeg-turbo & 2&77 & 52100 & 0.98 &58  &5.00\\
                                    &libpng        & 11&246 & 115380 & 0.73 &223  &3.05\\
                                    &libtiff       & 8&197 & 39137 & 0.74 &249 &2.80\\
                \midrule

                \multirowcell{9}{Data\\ Processing} &bsdiff              & 10&96         & 193024 &0.95  &247  &4.77\\
                                    &jsonnet             & 3&115       & 96601 &0.91   &78 &3.10 \\
                                    &snappy              & 3&34         & 102723 &0.82 &513  &3.95 \\
                                    &ledger              & 4&527        & 51645 &0.97  &108 &4.97 \\
                                    &leveldb             & 26&71       & 116955 &0.62   &482 &5.00 \\
                                    &cJson         & 4&78 & 21385 & 0.67 &389  &2.70\\
                                    &libmagic      & 4&18 & 3099 & 0.81 &163 &3.24\\
                                    &re2           & 3&70 & 20011 & 0.94 &129  &5.00\\
                                    &zlib          & 3&82 & 41311 & 0.53 &381 &1.74\\
                \midrule

                \multirowcell{4}{Hardware \\ \& System} &cpu\_features       & 13&42 & 46685 &0.92  &101  &4.73\\
                                    &cpuinfo             & 30&62      & 19118 &0.93   &47 &3.19 \\
                                    &puffin              & 8&80     & 12541 &0.88  &289 &2.08  \\
                                    &pthreadpool         & 4&54     & 24517 &0.90  &66   &3.05\\
                \midrule

                \multirowcell{8}{Web \& \\ Network} &libhtp              & 6&386        & 272602 &0.88   &149  &4.80\\
                                    &libphonenumber      & 25&377  & 78246 &0.80  &316  &4.26\\
                                    &minijail            & 7&162   & 188657 &0.72  &193  &2.76 \\
                                    &muduo               & 39&245  & 236759 &0.72  &275  &1.97  \\
                                    &uriparser           & 7&77    & 14583  &0.97  &188   &5.00\\
                                    &wabt                & 10&303  & 375987 &0.85  &496 &3.67 \\
                                    &c-ares        & 18&132 & 22650 & 0.82 &304  &3.91\\
                                    &libpcap       & 24&58 & 36858 & 0.49 &431 &2.04\\

                \midrule
                \midrule
                \multicolumn{2}{c|}{\textbf{Total}}          & 362 & 9860 &3068031 & - &4480 & 108.68\\
                \bottomrule
            \end{tabular}
        }
        
        \label{tab:stat for group and driver generation}
        \begin{tablenotes}
          \scriptsize
          \item  \textbf{I} = Implicit constraints;
          \textbf{E} = Explicit constraints;
        \end{tablenotes}
    \end{table}


%% file: figures/comparison_with_utopia.tex
    \begin{table}[t]
        \centering
        \caption{Comparison with Utopia.}
        \resizebox{.5\textwidth}{!}{
             \begin{tabular}{lll|ll|ll|ll}
                \toprule
                \multicolumn{3}{c|}{\textbf{Library}}                                & \multicolumn{2}{c|}{\textbf{Drivers}}                                                  & \multicolumn{2}{c|}{\textbf{Coverage (\%)}}                                                                                                & \multicolumn{2}{c}{\textbf{Bugs}}                    \\ 
                \midrule
                \textbf{Name} & \textbf{LoC} & \textbf{APIs} & \textbf{S} & \textbf{Utopia} & \textbf{S} & \textbf{Utopia} & \textbf{A} & \textbf{R} \\ 
                \midrule
                assimp & 356K       & 5055 & 106  & \textbf{250} & 15.22 & \textbf{36.3} & 7 & 7 \\
                bsdiff & 4K         & 137 & \textbf{247}  & 24 & \textbf{63.33} & 44.3 & 0 & 0 \\
                cpu\_features  & 6K & 36 & \textbf{101} & 4 & \textbf{43.39} & 9.6 & 0 & 0 \\
                cpuinfo & 423K      & 66 & \textbf{47}  & 6 & \textbf{30.91} & 16.2 & 0 & 0 \\
                jsonnet & 13K       & 98 & \textbf{78}  & 45 & \textbf{57} & 41.3 & 2 & 2 \\
                ledger & 51K        & 32 & \textbf{108} & 17 & \textbf{12.68} & 10.5 & 2 & 0 \\
                leveldb & 21K       & 397 & \textbf{482}  & 175 & \textbf{86.41} & 85.3 & 5 & 0 \\
                libaom & 363K       & 5065 & \textbf{117}  & 115 & \textbf{69.96} & 54.6 & 0 & 0 \\
                libhtp & 20K        & 386 & 149  & \textbf{336} & 76.05 & \textbf{78} & 0 & 0 \\
                libphonenumber & 53K  & 510 & \textbf{316} & 246 & 44.33 & \textbf{65.2} & 2 & 0 \\
                libvpx & 248K         & 1446 & \textbf{350}  & 40 & \textbf{49.45} & 34.3 & 0 & 0 \\
                minijail & 16K        & 162 & \textbf{193} & 159 & \textbf{42.77} & 39.6 & 2 & 0 \\
                muduo & 16K           & 359 & \textbf{275} &  5 & \textbf{13.19} & 11 & 3 & 0 \\
                openh264 & 92K        & 1523 & \textbf{152} & 113 & \textbf{50.97} & 34.3 & 1 & 0 \\
                pthreadpool & 12K     & 54 & 66 & \textbf{290} & \textbf{82.31} & 74.7 & 2 & 0 \\
                puffin & 5K           & 92 & \textbf{289} & 27 & \textbf{93.78} & 71.5 & 0 & 0 \\
                sfntly & 23K          & 897 & \textbf{174} & 16 & 24.46 & \textbf{49.3} & 0 & 0 \\
                snappy & 6K           & 46 & \textbf{513} & 14 & \textbf{93.68} & 79.5 & 2 & 0 \\
                uriparser & 8K        & 42 & \textbf{188} & 82 & 89.4 & \textbf{92.1} & 3 & 2 \\
                vowpal\_wabbit  & 81K & 1383 & 33 & \textbf{69} & \textbf{24.27} & 16.4 & 0 & 0 \\
                wabt & 47K            & 1034 & \textbf{496} & 80 & \textbf{39.27} & 26.3 & 1 & 1 \\
                \midrule
                \midrule
                \textbf{Total} & 1.8M   & 18820 &4880  & 2113 & 41.00 & - & 32 & 12 \\
                \bottomrule
            \end{tabular}
        }
        
        \label{tab:comparison with utopia}
        \begin{tablenotes}
          \tiny
          \item \textbf{A} = Number of total detected bugs;
          \textbf{R} = Number of reported bugs; 
          \item Since \utopia{} does not publicly disclose its total region coverage, we leave it empty.
          \item \textbf{S} = \toolname{}; 
        \end{tablenotes}
    \end{table}

%% file: figures/comparison_with_promptfuzz.tex
\begin{table*}[t]
    \centering
    \caption{Comparison with CKGFuzzer(CKG), PromptFuzz (PF) and OSS-Fuzz (OF).}
    \footnotesize
    \begin{tabular*}{.95\textwidth}{@{\extracolsep{\fill}} lcc|ccc|cc|cccc|cc}
        \toprule
        \multicolumn{3}{c|}{\textbf{Library}} & \multicolumn{3}{c|}{\textbf{Drivers}}     & \multicolumn{2}{c|}{\textbf{Query Costs (\$)}} & \multicolumn{4}{c|}{\textbf{Coverage (\%)}} & \multicolumn{2}{c}{\textbf{Bugs}} \\
        \midrule
        \textbf{Name} & \textbf{LoC} & \textbf{APIs} & \textbf{SD} & \textbf{CKG} & \textbf{PF} & \textbf{SD} & \textbf{PF} & \textbf{SD} & \textbf{CKG} & \textbf{PF} & \textbf{OF} & \textbf{A} & \textbf{R} \\
        \midrule
        c-ares        & 59K & 61  & \textbf{304} & 57  & 126 & 3.91 & 2.47 & \textbf{60.96} & 52.40 & 53.02 & 22.80 & 26 & 5 \\
        cJson         & 10K & 76  & \textbf{389} & 71  & 209 & 2.70 & 3.40 & \textbf{92.31} & 80.72 & 82.94 & 46.57 & 16 & 9 \\
        re2           & 28K & 70  & \textbf{129} & 68  & 101 & 5.00 & 3.30 & 73.42 & 67.00 & 64.61 & \textbf{78.94} & 1 & 0 \\
        lcms          & 45K & 286 & \textbf{488} & 247 & 402 & 4.83 & 14.04 & \textbf{52.69} & 44.11 & 42.49 & 34.62 & 0 & 0 \\
        libaom        & 530K & 47 & 175 & 45 & \textbf{237} & 5.00 & 4.11 & \textbf{46.38} & 22.29 & 25.62 & 18.01 & 0 & 0 \\
        libjpeg-turbo & 144K & 77 & 58 & 63 & \textbf{180} & 5.00 & 4.98 & 51.04 & 47.60 & 47.26 & \textbf{56.39} & 11 & 0 \\
        libmagic      & 33K & 18  & 163 & 17 & \textbf{217} & 3.24 & 2.41 & \textbf{75.74} & 37.99 & 63.67 & 62.74 & 11 & 5 \\
        libpcap       & 58K & 84  & \textbf{431} & 80 & 151 & 2.04 & 3.68 & \textbf{56.91} & 43.89 & 39.25 & 41.51 & 0 & 0 \\
        libpng        & 57K & 246 & 223 & 239 & \textbf{286} & 3.05 & 5.68 & \textbf{63.71} & 52.78 & 50.51 & 25.44 & 0 & 0 \\
        libtiff       & 108K & 195 & \textbf{249} & 167 & 153 & 2.80 & 4.02 & \textbf{55.59} & 43.70 & 52.43 & 40.60 & 19 & 1 \\
        libvpx        & 362K & 40 & 272 & 38 & \textbf{396} & 3.66 & 6.16 & \textbf{51.2} & 22.41 & 20.91 & 13.32 & 4 & 0 \\
        zlib          & 30K & 87  & \textbf{381} & 77 & 259 & 1.74 & 2.41 & \textbf{85.28} & 59.79 & 73.61 & 25.64 & 5 & 1 \\
        \midrule
        \midrule
        \textbf{Total} & 1.4M & 1287 & 3262 & 1288 & 2717 & 42.97 & 54.61 & \textbf{53.54} & 33.12 & 35.69 & 28.22 & 93 & 21 \\
        \bottomrule
    \end{tabular*}

    \label{tab:comparison with promptfuzz}
    \begin{tablenotes}
        \tiny
        \item
        \item \textbf{A} = Number of total detected bugs; \textbf{R} = Number of reported bugs;
        \item \textbf{SD} = \toolname{}; \textbf{CKG} = CKGFuzzer; \textbf{PF} = PromptFuzz; \textbf{OF} = OSS-Fuzz;
        \item \textbf{Total coverage} = (All covered branch) / (All branch) 
    \end{tablenotes}
\end{table*}

%% file: figures/ablation.tex
    \begin{table*}[t]
        \centering
        \caption{Component contributions of \toolname{}.}
        \resizebox{.95\textwidth}{!}{
             \begin{tabular}{l|llll|llll|llll}
                \toprule
                \textbf{Library} & \multicolumn{4}{c|}{\textbf{Drivers}}     & \multicolumn{4}{c|}{\textbf{Coverage (\%)}}   &  \multicolumn{4}{c}{\textbf{Early Crashes}}                \\ 
                \midrule
                \textbf{Name} & \textbf{w/o I} & \textbf{w/o \groupschedule{}} & \textbf{w/o \driverschedule{}} & \textbf{\toolname{}} & \textbf{w/o I} & \textbf{w/o \groupschedule{}} & \textbf{w/o \driverschedule{}} & \textbf{\toolname{}}  & \textbf{w/o I} & \textbf{w/o \groupschedule{}} & \textbf{w/o \driverschedule{}} & \textbf{\toolname{}}  \\
                \midrule
                assimp    & 73   & 110  &  172 & 106 & 4.02 & 7.60 & 6.02 & \textbf{15.22} &391 &196 &254 & \textbf{133} \\
                bsdiff    & 213   & 375  &  359 & 247 & 22.36 & 46.91 & 28.47 & \textbf{63.33} &72 &41 &44 & \textbf{28} \\
                cpu\_features    & 154   & 166  & 315  & 289 & 40.71 & 37.62 & 41.96 & \textbf{43.39} &290 &168 & 234 & \textbf{131} \\
                cpuinfo & 44   & 97  & 92  & 47 & 24.57 & 28.01 & 27.96 & \textbf{30.91} &329 &211 &241 & \textbf{179} \\
                jsonnet    & 39   &135   &  194 & 78 & 28.37 & 46.32 & 27.92 & \textbf{57.00} &314 &310 &344 & \textbf{194} \\
                ledger    & 47   & 169  &  141 & 108 & 6.48 & 10.15 & 6.57 & \textbf{12.68} &343 &83 &113 & \textbf{74} \\
                leveldb    & 353   & 787  &794  & 482 & 76.77 & 75.18 & 73.80 & \textbf{86.41} &857 &437 & 646& \textbf{319} \\
                libaom & 143   & 199  & 314 & 177 & 29.35 & 54.10 & 43.05 & \textbf{69.96} &316 &133 &168 & \textbf{35} \\
                libhtp    & 102   & 384  &  376 & 149 & 40.55 & 41.2 & 42.86 & \textbf{76.05} &833 &596 &707 & \textbf{308} \\
                libphonenumber    & 259   & 152  & 347 & 316 & 29.42 & 34.62 & 32.42 & \textbf{44.33} &216 &\textbf{163} &169 & 195 \\
                libvpx    & 300   & 400  & 490 & 350 & 48.2 & 47.51 & 48.95 & \textbf{49.45} &497 &376 & 421 & \textbf{354} \\
                minijail & 119   & 290  & 307  & 193 & 34.43 & 40.17 & 37.40 & \textbf{42.77} &681 &650 &715 & \textbf{586} \\
                muduo    & 257   & 384  & 438 & 275 & 6.11 & 9.01 & 7.65 & \textbf{13.19} &376 &269 &274 & \textbf{244} \\
                openh264    & 83   & 157  & 171 & 152 & 24.39 & 47.43 & 30.64 & \textbf{50.97} &676 &190 &223 & \textbf{147} \\
                pthreadpool    & 87   & 144  & 175  & 66 & 81.15 & 82.05 & 76.22 & \textbf{82.31} &467 &411 & 557& \textbf{375} \\
                puffin & 195   & 280  & 438 & 289 & 80.43 & 73.58 & 93.36 & \textbf{93.78} &134 &79 &122 & \textbf{68} \\
                sfntly    & 85   & 127  &  213 & 174 & 10.71 & 11.71 & 11.02 & \textbf{24.46} &408 &415 &557 & \textbf{270} \\
                snappy    & 316   & 356  & 422 & 513 & 89.92 & 92.34 & 82.21 & \textbf{93.68} &433 &\textbf{277} &313 & 387 \\
                uriparser    & 140   & 116  & 130  & 188 & 81.04 & 81.69 & 80.96 & \textbf{89.40} &154 &151 & 110& \textbf{91} \\
                vowpal\_wabbit & 26   & 36  & 45  & 33 & 11.61 & 10.28 & 10.18 & \textbf{24.47} &64 &\textbf{45} &58 & 47 \\
                wabt    & 426   & 533  & 534 & 496 & 20.24 & 39.20 & 22.37 & \textbf{39.27} &387 &412 &442 & \textbf{354} \\
                c-ares    & 293   & 358  & 375 & 304 & 48.30 & 56.12 & 51.23 & \textbf{60.96} &389 &366 &360 & \textbf{319} \\
                cjson    & 243   & 432  & 442  & 389 & 77.91 & 89.96 & 83.45 & \textbf{92.31} &532 &255 & 289 & \textbf{178} \\
                re2 & 72   & 107  & 114  & 129 & 25.91 & 57.44 & 51.50 & \textbf{73.42} &218 &219 &352 & \textbf{145} \\
                lcm3    & 395   & 520  & 530 & 488 & 40.76 & 45.44 & 44.19 & \textbf{52.69} &603 &578 &543 & \textbf{430} \\
                libaom    & 177   & 231  & 327 & 175 & 27.92 & 43.34 & 42.27 & \textbf{46.38} &284 &93 &103 & \textbf{41} \\
                libjpeg-turbo    & 54   & 124  & 168  & 58 & 22.14 & 26.12 & 25.64 & \textbf{51.04} &433 &278 & 316 & \textbf{291} \\
                libmagic & 138   & 320  & 360  & 163 & 22.09 & 43.92 & 43.66 & \textbf{75.74} &699 &667 &512 & \textbf{451} \\
                libpcap    & 361   & 430  & 513 & 431 & 51.52 & 54.94 & 50.49 & \textbf{56.91} &776 &773 &665 & \textbf{511} \\
                libpng    & 161   & 301  & 313 & 223 & 19.36 & 42.39 & 30.35 & \textbf{63.71} &475 &490 &591 & \textbf{344} \\
                libtiff    & 240   & 286  & 302  & 249 & 25.91 & 43.42 & 38.32 & \textbf{55.59} &309 &194 & 276 & \textbf{147} \\
                libvpx & 206   & 301  & 327 & 272 & 33.52 & 49.12 & 43.85 & \textbf{51.20} &606 &350 &430 & \textbf{324} \\
                zlib    & 345   & 364  &  433 & 381 & 83.33 & 84.90 & 84.49 & \textbf{85.28} &465 &495 &664 & \textbf{421} \\                
                \midrule
                \midrule
                \textbf{Total} & 6146  & 9171 & 10671  & 7742 & 38.59 & 47.09 & 43.38 & \textbf{56.61} &14036 &10407 &11597 & \textbf{8121} \\
                \bottomrule
            \end{tabular}
        }
        \label{tab:ablation}
        \begin{tablenotes}
          \tiny
          \item
          \textbf{w/o} = without; \textbf{I} = Implicit constraints; 
        \end{tablenotes}
    \end{table*}

%% file: sections/5_1_rq1.tex
\subsection{Effectiveness of Driver Generation}
As shown in \autoref{tab:stat for group and driver generation},
\toolname{} utilizes the LLM to extract overall 362 implicit constraints and 9,860 explicit constraints.
Consequently, it solves them to generate 3,068,031 API groups and synthesize 4,480 drivers in our experiments.

\boldparagraph{Constraints.} 
In general, more explicit constraints lead to more groups, while implicit constraints reduce the number of groups. In the category \textit{Web \& Network}, \toolname{} extracts the most implicit constraints on average (17), since the libraries emphasize API composition to construct an application. While the category \textit{Data Processing} has the fewest implicit constraints on average (6.67), the libraries tend to generalize the APIs instead of imposing specialized semantic data formats.

We sample ten implicit constraints from each library to evaluate generation accuracy; for libraries with fewer than ten implicit constraints, all their constraints are included in the analysis.
To determine the correctness of each constraint, we review the function definitions and comments of the APIs associated with it. An implied constraint is classified as a false positive (FP) if its correct usage results in a conflict; otherwise, it is considered a true positive. Conversely, constraints with conflict type but whose correct type is imply are treated in the same manner. The analysis shows that 31 out of 228 constraints are classified as FPs. Notably, 15 of these 31 FPs are due to the APIs of the constraints not being functions. These results demonstrate that LLMs can effectively generate implicit constraints to help library fuzzing techniques.

\boldparagraph{Driver Generation.}
Within our generation framework, the stillborn rate ranges from 0.49 to 0.98. 
Notably, the number of drivers is not linearly related to the number of groups, as some generated drivers might be filtered due to early-stage crashes. 
For instance, the library openh264 has 453,012 groups, while we only obtained 152 drivers. However, \toolname{} generates 488 drivers for lcms with 68,478 groups. 
The number of drivers depends on the complexity of API design; in short, the fewer parameters required by the APIs, the easier it is to generate drivers. 

\input{figures/failure_analysis_compilation_error}

To understand the causes of driver generation failures, we automatically classify driver compilation errors using a string-based pattern matching approach (details in \autoref{sec: Failure Analysis Patterns of Driver Generation}). A total of 18,154 compilation errors are analyzed, as shown in \autoref{fig:failure analysis compilation error}. These errors are categorized into six groups, four of which are derived from previous work~\cite{10.1145/3650212.3680355}, while the remaining two are specific to the benchmark of our study:

\begin{itemize}
    \item \textit{G1 - Corrupted Code (401)}: Drivers lack complete code due to issues such as missing implementations, undefined types, missing terminator tokens (e.g., semicolons), or mismatched brackets.
    \item \textit{G2 - Language Basics Violation (6720)}: Code violates fundamental language rules, such as redefinitions, improper access controls (e.g., \inlinecpp{private}), or ambiguous name resolutions.
    \item \textit{G3 - Non-Existing Identifier (6828)}: Code reference undefined symbols, including functions, member methods, constructors, and variables.
    \item \textit{G4 - Type Error (2189)}: Errors occur due to mismatched function arguments, invalid type casts, incorrect usage of left and right values, or operations like arithmetic on pointers.
    \item \textit{G5 - Token Limitation (820)}: One major cause occurs during the linking stage, where numerous unresolved references result in compilation error messages that exceed GPT's token limit. Refining these errors to address this limitation is deferred to future work.
    \item \textit{G6 - Out of Space (1196)}: Certain libraries (e.g., libjpeg-turbo) provide APIs that interact with the local file system, including file creation. This results in drivers consuming excessive storage in the fuzzing environment, ultimately leading to a lack of available space on the machine.
\end{itemize}

\boldparagraph{Generation Time.}
In our experiments, we set the query cost limitation to \$5 per library. As indicated in \autoref{tab:stat for group and driver generation}, five libraries (libaom, jpeg, leveldb, re2, and uriparser) reached the limit after fuzzing began at 17h, 14h, 12h, 8h, and 12h, respectively. 
Since these libraries have a lower SR, \toolname{} can easily generate valid drivers for them and query the LLM more frequently than others. 
Thus, it quickly runs out of the query cost limit for these five libraries. 
For the others, \toolname{} spends more time constructing prompts, resulting in lower query costs.

\boldparagraph{Coverage of Drivers.}
\autoref{fig:coverage} displays the overall accumulated code coverage achieved by the \toolname{}'s generated fuzz drivers, and
a continuous increase over time can be observed across all the libraries. 
In some cases (e.g., openh264, pthreadpool, libmagic), \toolname{} can discover crucial drivers to break through the coverage limitation, resulting in multiple steep increases during the middle and later fuzzing periods. 
For the library sfntly, \toolname{} initially achieves zero coverage since it fails to generate effective drivers. 
However, the scheduling components continuously attempt to select possible API groups and drivers, eventually achieving 24.5\% region coverage.

\begin{tcolorbox}[size=title]
{\textbf{Answer to Effectiveness of Driver Generation:} 
\toolname{} successfully extracts implicit and explicit constraints and uses them to generate API groups. In addition, it can continually generate valid and rational drivers based on the API groups throughout the entire fuzzing campaign.
}
\end{tcolorbox}

%% file: figures/failure_analysis_compilation_error.tex
\begin{figure}[t]
    \centering
        \includegraphics[width=0.45\textwidth]{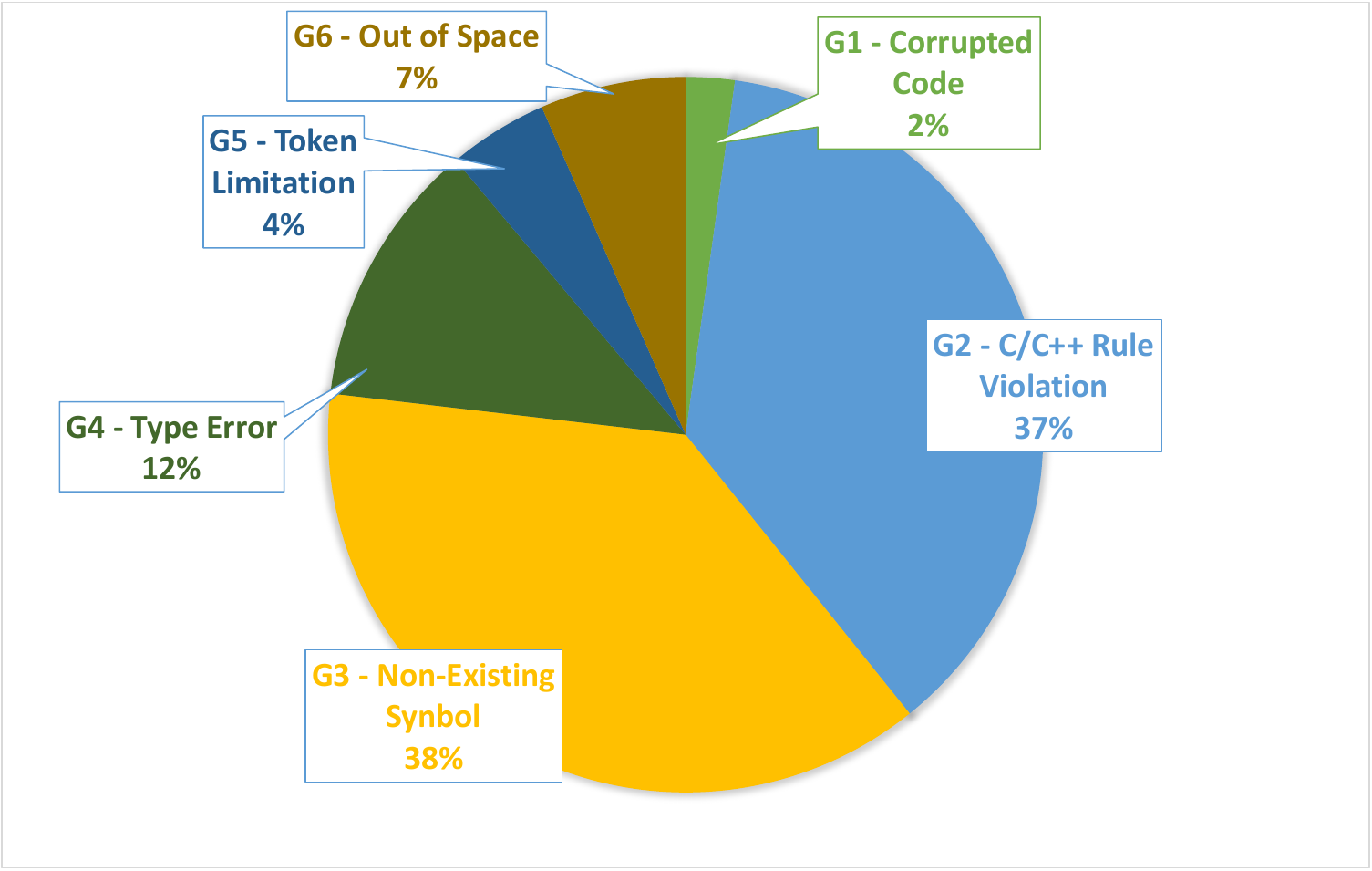}
        \caption{Failure Taxonomy of Driver Generation.}
    \label{fig:failure analysis compilation error}

\end{figure}

%% file: sections/5_2_rq2.tex
\subsection{Advancement of \toolname{}}
To evaluate how \toolname{} performs better than baseline techniques, 
we compare the coverage of libraries against \utopia{}, \ckgfuzzer{} \promptfuzz{}, and \ossfuzz{}. 
For the comparison with \utopia{}, we conducted our experiments on 21 out of 25 libraries since 4 libraries are uncompilable in \utopia{}. 
Meanwhile, we evaluated \toolname{} on 12 out of 14 libraries from \promptfuzz{} \ckgfuzzer{} and \ossfuzz{} since 2 of them are unproducible as well.


\boldparagraph{Comparison with \utopia{}.} 
\utopia{} is a unit-test-based driver synthesis technique that uses static analysis. 
The evaluation results are shown in \autoref{tab:comparison with utopia}. 
Overall, \toolname{} generates twice as many fuzz drivers as \utopia{} in total and outperforms it in 17 out of 21 libraries.
For the remaining 4 libraries where \toolname{} generates fewer drivers, \textit{assimp} and \textit{vowpal\_wabbit} each have over 1000 extracted APIs. 
The large number of APIs results in poor performance in constraint analysis and makes it difficult to hit crucial API groups. 
For \textit{libhtp}, a network library, a typical calling sequence involves many steps such as registering, opening, handshaking, and closing. 
Thus, the complexity of the API usage scenarios leads to a low driver generation rate. 
On the other hand, some APIs in the \textit{pthreadpool} library require over 10 arguments, which the LLM also cannot process well.

Compared in terms of region coverage, \toolname{} achieves higher region coverage than \utopia{} on most libraries (16/21).
In general, more valid drivers evidently lead to higher region coverage. 
Nevertheless, the libraries \textit{libphonenumber}, \textit{sfntly}, and \textit{uriparser} showed opposite results. 
These three libraries require structured inputs, which \toolname{} does not aim to handle. 
Therefore, \utopia{} performs well by leveraging unit tests to construct structured inputs, while \toolname{}'s drivers generate random inputs that are rejected early.

\boldparagraph{Comparison with \ckgfuzzer{}, \promptfuzz{} and \ossfuzz{}.} 
\promptfuzz{} and \ckgfuzzer{} represent state-of-the-art LLM-based library fuzzing techniques. As shown in \autoref{tab:comparison with promptfuzz}, \toolname{} achieves significantly higher overall branch coverage (53.54\%) compared to \ckgfuzzer{} (33.12\%) and \promptfuzz{} (35.69\%). Notably, \toolname{} outperforms both across all evaluated libraries, despite spending less time on driver generation and fuzzing. Regarding driver quantity, \toolname{} generates more drivers than \promptfuzz{} in 7 out of 12 libraries, and consistently surpasses \ckgfuzzer{}. These results highlight \toolname{}’s ability to synthesize higher-quality drivers and schedule them more effectively.

As a manually crafted fuzz driver project, \ossfuzz{} has benefited the open-source community for years. 
Compared with \ossfuzz{}, \toolname{} also achieves higher total branch coverage (53.54\%) than \ossfuzz{} (28.22\%). 
Given the human workload required by \ossfuzz{}, \toolname{} not only generates better drivers but also requires less effort.

\begin{tcolorbox}[size=title]
{\textbf{Answer to Advancement:} 
\toolname{} outperforms all the baseline techniques overall, requiring less time, lower LLM token fees, and less human effort. It can generate better drivers and schedule them efficiently to achieve optimal code coverage.
}
\end{tcolorbox}

%% file: figures/ablation_coverage.tex




\begin{figure*}[t]

    \begin{subfigure}{\textwidth}
            \includegraphics[width=1\textwidth]{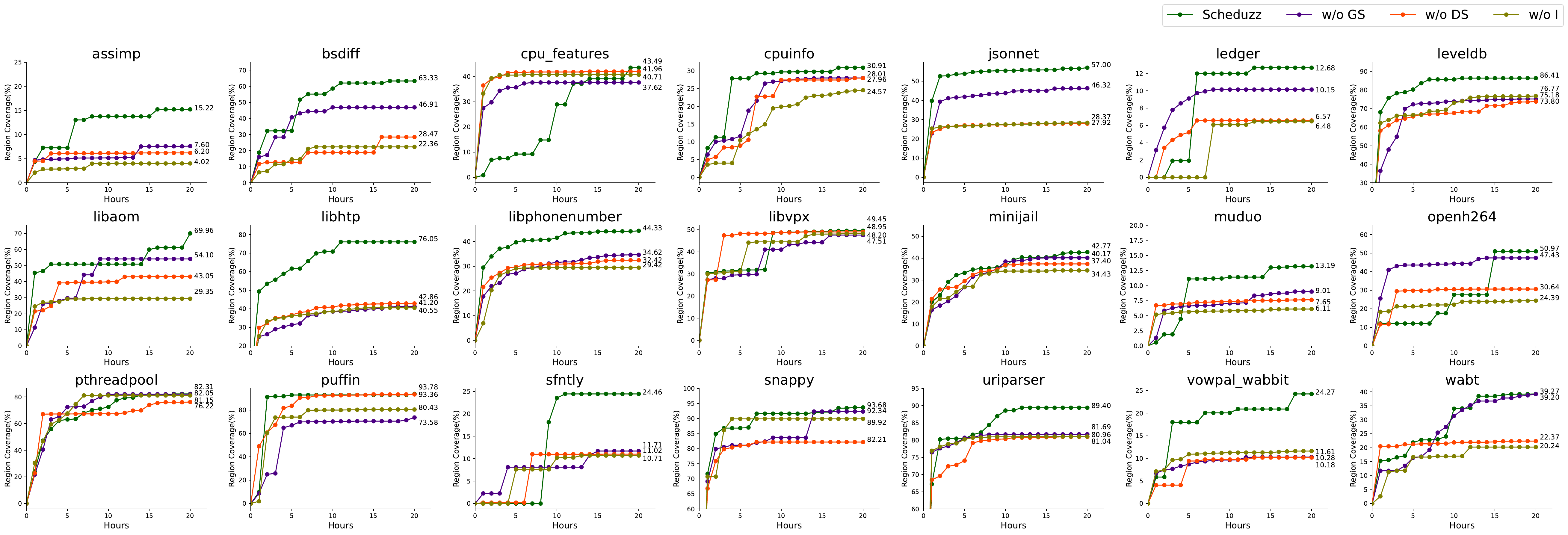}
            \caption{The region coverage on the libraries from \utopia{}.}
            \label{fig:ablation ut}
    \end{subfigure}

    \begin{subfigure}{\textwidth}
            \includegraphics[width=1\textwidth]{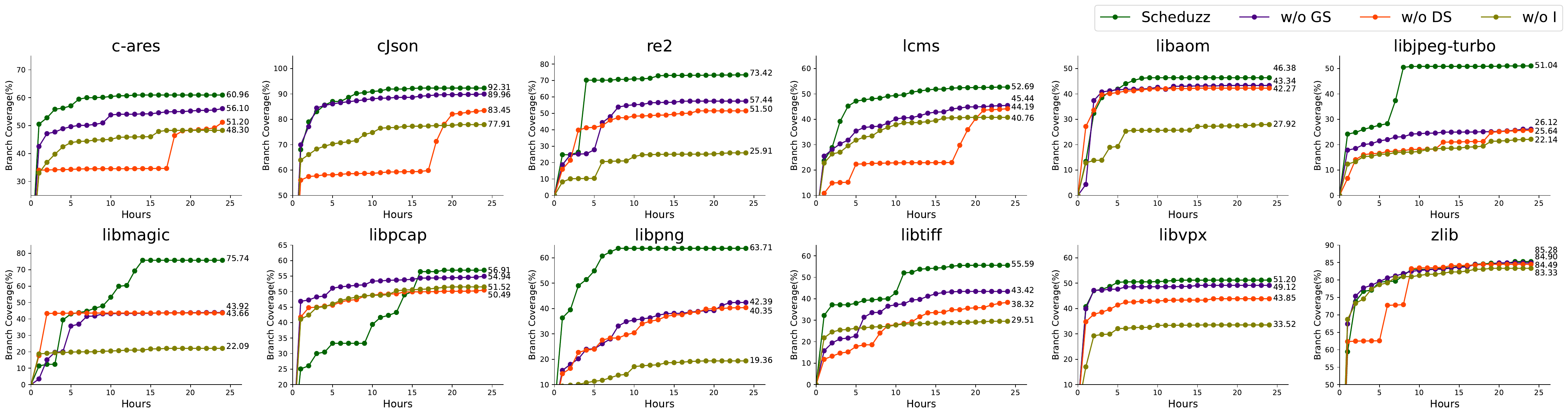}
            \caption{The branch coverage on the libraries from \promptfuzz{}.}
            \label{fig:ablation pf}
    \end{subfigure}

    \caption{Coverage of \toolname{} with each component disabled.}
    \label{fig:ablation_coverage}

\end{figure*}

%% file: sections/5_3_rq3.tex
\subsection{Necessity of Each Component}
We conduct \toolname{} on the benchmarks with each component individually disabled. 
Specifically, we remove the implicit constraints in the \textit{Group Generation} (\groupgen{}) component, leverage random selection in \groupschedule{}, and utilize \textit{Round-robin}~\cite{miao2016fundamentals} scheduling for \driverschedule{}.

Overall, \groupgen{} is the most impactful component in our experiments, as shown in \autoref{tab:ablation} and \autoref{fig:ablation_coverage}. Its absence leads to the worst performance in terms of average coverage (38.59\%). The subsequent components are \driverschedule{} and \groupschedule{}, which achieve averages of 43.38\% and 47.09\%, respectively. These results explicitly demonstrate the importance and contributions of both API constraints and the dual scheduling framework in enhancing the overall performance.
We evaluate the rationality of drivers by counting the number of early crashes in the \driverschedule{} filter.
More early crashes indicate that more generated drivers can be compiled while still consisting of irrational API calling sequences.
In \autoref{tab:ablation}, \toolname{} detect 1.73 times more early crashes in total when the implicit constraints are removed, leading to 1596 fewer generated useless fuzz drivers. The experimental result shows that constraints can appropriately model API rationality and improve the fuzzing efficiency.
Moreover, not only can implicit constraints directly contribute to the rationality of drivers, but \groupschedule{} and \driverschedule{} also benefit it indirectly. 
Without these two schedulers, \toolname{} raises more than 2286 and 3476 early crashes, respectively, because both schedulers can compute lower scores for plausible early crash API groups and drivers, reducing the possibility of selecting irrational ones.

\begin{tcolorbox}[size=title]
{\textbf{Answer to Necessity:} 
Each component contributes to improving \toolname{}'s performance in driver generation and achieving code coverage. Among these components, implicit constraint outperforms the others.
}
\end{tcolorbox}

%% file: sections/5_4_rq4.tex
\subsection{Detected Bugs and Drivers}
To demonstrate the capability of driver generation, 
we summarize some typical confirmed defects and its corresponding fuzz drivers as follows.

\boldparagraph{Rationality.}
In \autoref{fig:bug1}, the driver uses multiple APIs and leverages implicit constraints to free all allocated memory at the driver's end. 
If the driver fails to free any memory, the fuzzing campaign will terminate immediately because the sanitizer detects the memory leak error. 
Our design helps eliminate false positives caused by irrational API usage. This driver detects integer and buffer overflow defects in \inlinecpp{URI_FUNC(ComposeQueryEx)}, for which two CVEs have been assigned.

\boldparagraph{API Missing.}
\toolname{} generates a driver that detects an out-of-memory error, as shown in \autoref{fig:bug2}. 
The error occurs because \inlinecpp{aiGetPredefinedLogStream} pushes an object into a global vector without ever releasing it. 
\toolname{} fails to analyze the implicit constraints because there is no corresponding release API for it. 
Therefore, the implicit constraints can detect the errors caused by improper API design of library developers.
Surprisingly, \toolname{} can help library developers complete the missing API by constructing edge case drivers.

\begin{tcolorbox}[size=title]
{\textbf{Answer to Case Study:} 
\toolname{} can generate drivers with rational API usages and detect the regular defects. Moreover, it can find the improper API design by leveraging implicit constraints, which are beyond the detection capabilities of existing techniques.
}
\end{tcolorbox}

%% file: figures/bug1.tex
\begin{figure}[t]
    \centering

    \includegraphics[width=.5\textwidth]{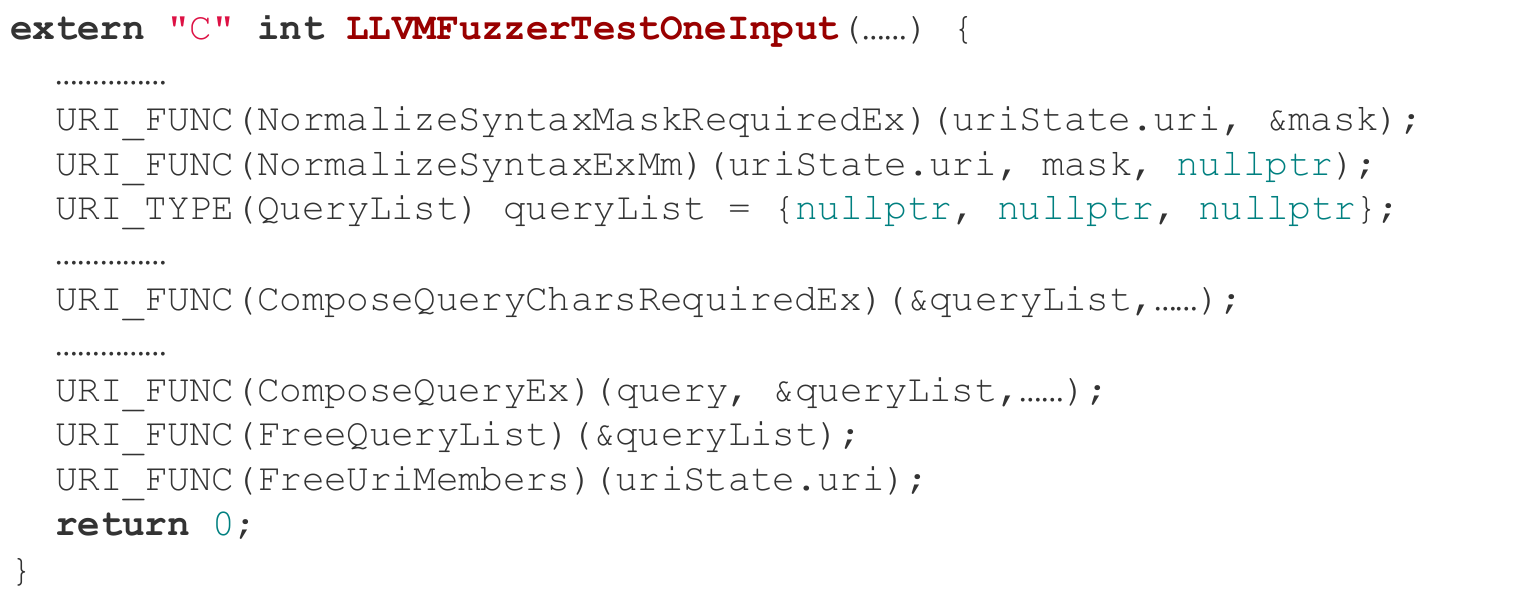}
    
    \caption{The driver detects integer and buffer overflow vulnerability (CVE-2024-34402, CVE-2024-34403).}
    \label{fig:bug1}

\end{figure}

%% file: figures/bug2.tex
\begin{figure}[t]
    \centering

    \includegraphics[width=.5\textwidth]{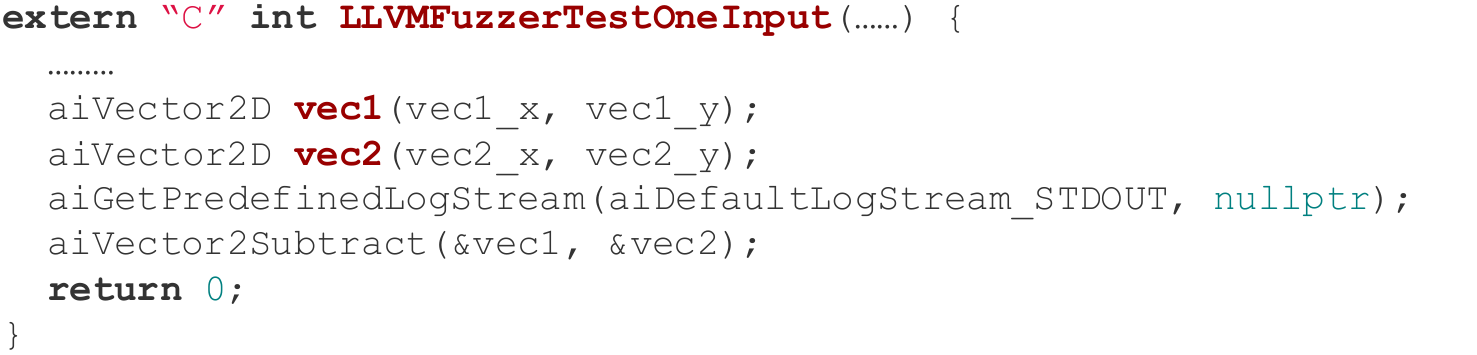}
    
    \caption{The driver detects a missing API.}
    \label{fig:bug2}
    \vspace{-5mm}

\end{figure}

%% file: sections/6_0_related_work.tex
\section{Related Work}
\boldparagraph{Fuzz driver generation.}
In recent years, many approaches for fuzz driver generation have been proposed. These approaches can be roughly categorized into two types: non-learning-based and learning-based.

In non-learning-based approaches~\cite{babic2019fudge,ispoglou2020fuzzgen,jiang2021rulf,jeong2023utopia,zhang2021apicraft,jung2021winnie,chen2023hopper,10.5555/3620237.3620398}, researchers obtain drivers by leveraging static and dynamic analyses to mutate initial seeds. Fudge~\cite{babic2019fudge}, FuzzGen~\cite{ispoglou2020fuzzgen}, RULF~\cite{jiang2021rulf}, and Utopia~\cite{jeong2023utopia} perform static analysis to extract calling sequences from existing code. Specifically, FuzzGen and RULF construct an API dependence graph and traverse it to synthesize the calling sequences, while Utopia mutates unit test cases and converts them into libFuzzer-style drivers. On the other hand, some dynamic analysis-based approaches~\cite{zhang2021apicraft, jung2021winnie} analyze execution traces to automatically create fuzz drivers. Hopper~\cite{chen2023hopper} synthesizes fuzz drivers by fuzzing an interpreter to obtain calling sequences from libraries.

To eliminate reliance on initial seeds and sample code, several learning-based approaches~\cite{deng2023large, 10.1145/3597503.3623343, lyu2023prompt} have been proposed to generate drivers automatically. 
TitanFuzz~\cite{deng2023large} and PromptFuzz~\cite{lyu2023prompt} use LLMs to generate fuzz drivers through filling and mutation.
FuzzGPT~\cite{10.1145/3597503.3623343} fully automated leverages historical information to generate deep learning library drivers via the intrinsic capabilities of LLMs.

\boldparagraph{LLM-based software testing.}
LLMs are increasingly being utilized in software testing and it has been applied for test inputs generation and defect detection mainly. 

In the approaches of generating test inputs based on LLMs~\cite{zhang2024llamafuzz,li2024acecoder,meng2024large,vikram2023can,xia2024fuzz4all,schafer2023empirical,xie2023chatunitest}, LLAMAFUZZ~\cite{zhang2024llamafuzz} fine-tune LLMs using historical test input datasets to generate a large number of inputs that meet testing requirements. Acecoder~\cite{li2024acecoder}, ChatAFL~\cite{meng2024large}, PBT-GPT~\cite{vikram2023can}, Fuzz4All~\cite{xia2024fuzz4all} utilizes zero-shot learning, few-shots learning, chain of thought, and automatic prompt methods in prompt engineering to generate test inputs without fine-tuning LLMs. Meanwhile, TESTPILOT\cite{schafer2023empirical} and ChatUniTest~\cite{xie2023chatunitest} introduce feedback frameworks to provide feedback to the model for modifying erroneous test inputs generated by LLMs, and continuously iterate to obtain better generation results.

In the defect detection process, some methods such as DIVAS~\cite{paria2023divas} use LLMs to learn vulnerability features to generate test oracles and use them for defect detection. On the other hands, some methods~\cite{kang2023explainable,kang2023preliminary} such as AutoSD~\cite{kang2023explainable} and AutoFL~\cite{kang2023preliminary} use LLMs to directly analyze code snippets, predict and locate potential vulnerabilities.


%% file: sections/7_0_discussion_limitation.tex
\section{Discussion and Limitation}
\boldparagraph{Implicit and Explicit Constraints.}
First, \toolname{} aims to analyze the rational usage of third-party libraries by leveraging the LLM. After analyzing, \toolname{} obtains the implicit constraints, which encode the relationships between APIs. 
However, due to hallucination of LLMs, it's difficult to verify whether the implicit constraints appropriately describe rational usage. 
To make matters worse, we don't know the number of constraints in a library or when the analysis is saturated.
Therefore, in future work, we will explore techniques such as Retrieval-Augmented Generation (RAG), the utilization of knowledge graphs, and other hallucination-reducing methods~\cite{tonmoy2024comprehensive} to refine implicit constraints.
Second, the explicit constraints broadly consider all parameters as inputs and outputs, which can result in false positives during the generation of rational fuzz drivers.
To enhance explicit constraints with input and output information, 
we plan to explore LLM-based API input and output analysis in future work.


\boldparagraph{Large Library in Practice.}
A few libraries contain hundreds of APIs, which can create significant pressure on constraint solvers. However, as mentioned earlier, since a unit of functionality typically requires only a limited set of APIs, this prior knowledge can assist constraint solvers in efficiently handling large-scale libraries.
Additionally, large-scale libraries often include complex APIs with numerous parameters, increasing the likelihood of hallucinations from LLMs. One specific potential solution is to decompose complex APIs into multiple simpler virtual ones and use our devised constraint "imply" to establish relationships between them.
Moreover, the LLM community has proposed various hallucination-reducing methods~\cite{tonmoy2024comprehensive}, which we can leverage to further enhance robustness.

%% file: sections/8_0_conclusion.tex
\section{Conclusion}
In this paper, we develop \toolname{}, which automatically generates fuzz drivers for widely used C/C++ libraries without human intervention. 
In \toolname{}, we devise two types of API constraints to model valid and rational API usage. 
Furthermore, we leverage the LLM to extract these constraints, generate numerous fuzz drivers, and schedule them to fuzz libraries efficiently. 
The experimental results show that \toolname{} achieves higher code coverage compared to a handcrafted fuzz driver project and two state-of-the-art techniques. In a short-term fuzzing campaign, \toolname{} found 3 CVEs in well-tested libraries.

%% file: sections/11_appendix.tex
\section{Failure Analysis}
\label{sec: Failure Analysis Patterns of Driver Generation}
\input{appendix/figures/patterns_for_driver_generation_-_g1_g2}

\boldparagraph{Failure Analysis Patterns of Driver Generation.}
We employ 114 string-based patterns to classify failures in driver generation, as detailed in \autoref{tab:patterns}. For instance, a failure is categorized as \textit{G1 - Corrupted Code} if the build error message contains a string such as \inlinecpp{error: expected ')'} from the pattern G1, as shown in \autoref{tab:patterns}.

\input{appendix/figures/repeat}

\boldparagraph{Failure Rate.}
As illustrated in \autoref{fig:Failure Rate of Repeat Times}, repeated queries significantly reduce the failure rate of driver generation for both missing APIs and compilation errors.

In the first query attempt, we leverage the dual-scheduler framework to select a group for driver generation and compilation. For subsequent attempts (from the second to the tenth), we focus on the drivers filtered out during the previous generation process. Using the same approach, we construct tailored prompts to provide context for the LLMs.
The experimental results identify a fixed point where further repetitions yield diminishing returns. Based on these findings, we select four as the maximum number of query attempts to achieve an optimal balance between generation success and token cost.

\section{Regular Expression of Retriever.}
\label{sec: Regular Expression of Retriever}
We implement the information retriever in Python. The core approach involves using regular expressions (regex) to extract key information from compiler error messages. The specific patterns used in \toolname{} are detailed as follows:
\begin{lstlisting}[language=C,showstringspaces=false ]
r"error: no matching function for call to '([^']*)'"
r"error: use of undeclared identifier '([^']*)'"
r"error: use of undeclared identifier ([^']*); did you mean '([^']*)'\?"
r"error: assigning to '([^']*)'(?: \(aka '[^']*'\))? from incompatible type '[^']*'(?: \(aka '[^']*'\))?"
r"error: unknown type name '([^']*)'"
r"error: no member named '[^']*' in '([^']*)'"
r"error: field designator '[^']*' does not refer to any field in type '([^']*)'(?: \(aka '[^']*'\))?"
\end{lstlisting}

%% file: appendix/figures/patterns_for_driver_generation_-_g1_g2.tex
\begin{table*}[t]
\scriptsize
\resizebox{.79\textwidth}{!}{
            \begin{tabular}{l|l}
            \toprule
            \textbf{Pattern for G1 - Corrupted Code} & \textbf{Pattern for G3 - Non-Existing Identifier } \\ 
            \midrule
            is an abstract class & no matching function for call to \\c
            error: no viable conversion from & error: use of undeclared identifier \\
            error: variable has incomplete type & undefined reference to \\
            error: expected '\}' & error: no member named \\
            error: expected ')' & no matching constructor for initialization \\
            error: expected ';' after expression & error: no matching member \\
            error: expected expression & error: field designator \\
            \cmidrule[\heavyrulewidth]{2-2}
            error: expected '>' & \textbf{Pattern for G4 - Type Error} \\
            \cmidrule[\heavyrulewidth]{2-2}
            error: extraneous closing brace &error: no type named\\
            named in nested name specifier &error: unknown type name\\
            error: a type specifier is required for all declarations& invalid operands to binary expression\\
            error: C++ requires a type specifier &error: unexpected type name\\
            error: expected unqualified-id& error: member reference base type\\
            error: extraneous ')' before ';' &error: cannot initialize a\\
            error: variable declaration in condition cannot have a parenthesized initializer& error: reinterpret\_cast from \\
            does not name a template but is followed by template arguments& error: member access into \\
            error: templates must have C++ linkage& error: incompatible integer to pointer conversion\\
            tag to refer to type & from incompatible type\\
            does not refer to a value & error: cast from pointer to smaller type\\ 
            \cmidrule[\heavyrulewidth]{1-1}
            \textbf{Pattern for G2 - Language Basics Violation } &error: incompatible pointer types\\ 
            \cmidrule[\heavyrulewidth]{1-1}
            multiple definition of &is not a function or function pointe\\
            error: calling a protected constructor &error: non-constant-expression cannot be narrowed from type\\
            error: attempt to use a deleted function& error: invalid use of incomplete type\\
            error: overload resolution selected deleted operator& error: cannot cast from type\\
            cannot be implicitly captured in a lambda with no capture-default specified &has incompatible initializer of type\\
            error: redefinition of &error: const\_cast from\\
            discards qualifiers&error: static\_cast from \\
            error: invalid application of& cannot be narrowed to\\
            error: cannot jump from this goto statement to its label &error: invalid argument type \\
            is ambiguous &error: incompatible pointer to integer conversion\\
            error: function definition is not allowed here &error: too few arguments to function call\\
            has a different language linkage &error: invalid range expression of type\\
            error: typedef redefinition with different types &s not a pointer; did you mean to use\\
            error: illegal initializer & error: conflicting types\\
            error: excess elements in scalar initializer& error: non-const lvalue reference to typ \\
            error: call to non-static member function &error: no matching conversion for\\
            is a private member of &error: too many arguments to function call \\
            error: multiple overloads of &error: arithmetic on a pointer\\
            is a protected member of &  error: comparison between \\
            error: call to implicitly-deleted default constructor of &  error: C-style cast from \\
            error: reference to non-static member function &error: incompatible operand types\\
            error: expression is not assignable  &could not bind to an lvalue of type \\
            error: call to implicitly-deleted &error: cannot take the address of an rvalue of type \\
            could not bind to an rvalue of type &error: too many arguments provided \\
            is neither visible in the template definition nor found by argument-dependent lookup &error: cannot initialize an \\
            error: ambiguous conversion &is not assignable\\
            error: no viable overloaded &error: functional-style cast\\
            error: calling a private &must match previous return type\\
            error: allocating an object of abstract class type &error: cannot compile this lambda conversion to variadic function yet\\
            is a pointer; did you mean to use &is not contextually convertible \\
            error: only virtual member functions &cannot be referenced with a struct specifier \\
            error: cannot jump from &error: cannot convert\\
            error: cannot delete &error: indirection requires pointer operand\\
            does not provide a call operator &error: functions that differ only in their return type cannot be overloaded\\
            error: excess elements in struct initializer &\\
            error: taking the address of a temporary objec &\\
            used in function with fixed args& \\
            error: reference to overloaded function could not be resolved& \\
            variables must have global storage& \\
            conflicts with typedef of the same name &\\
            error: call to deleted& \\
            error: cannot create a non-constant pointer to member function &\\
            \bottomrule 
            \end{tabular}
}
    \caption{Failure Analysis Patterns.}
    \label{tab:patterns}
    \vspace{-8mm}
\end{table*}

%% file: appendix/figures/repeat.tex
\begin{figure}[h]
        \centering
        \includegraphics[width=0.4\textwidth]{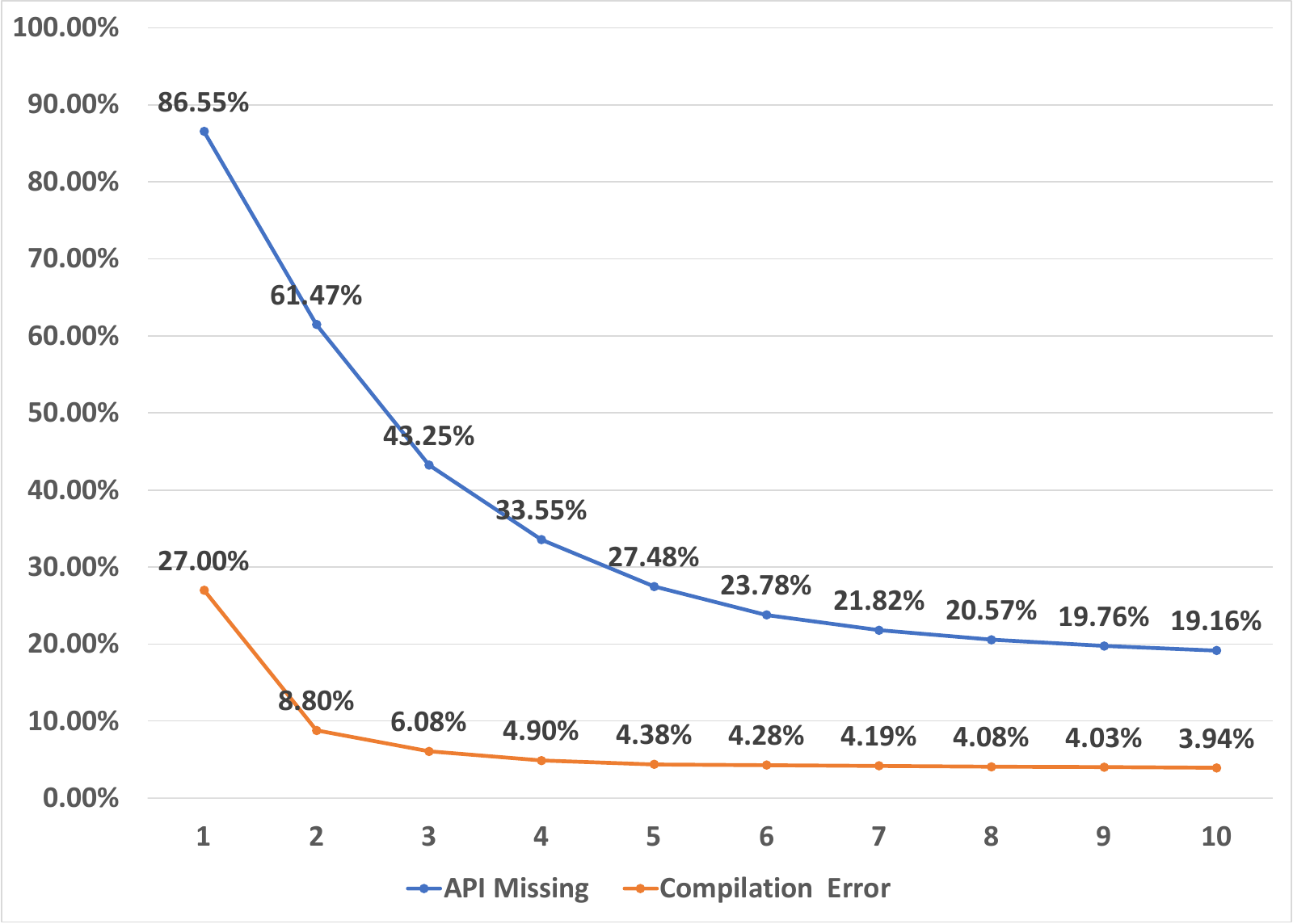}
        \caption{Failure Rate vs. Repeat Query Times. The x-axis represents the number of repeat query attempts, while the y-axis shows the corresponding failure rate.}
        \label{fig:Failure Rate of Repeat Times}
\end{figure}

%% file: main.bbl

\begin{thebibliography}{37}


\ifx \showCODEN    \undefined \def \showCODEN     #1{\unskip}     \fi
\ifx \showDOI      \undefined \def \showDOI       #1{#1}\fi
\ifx \showISBNx    \undefined \def \showISBNx     #1{\unskip}     \fi
\ifx \showISBNxiii \undefined \def \showISBNxiii  #1{\unskip}     \fi
\ifx \showISSN     \undefined \def \showISSN      #1{\unskip}     \fi
\ifx \showLCCN     \undefined \def \showLCCN      #1{\unskip}     \fi
\ifx \shownote     \undefined \def \shownote      #1{#1}          \fi
\ifx \showarticletitle \undefined \def \showarticletitle #1{#1}   \fi
\ifx \showURL      \undefined \def \showURL       {\relax}        \fi
\providecommand\bibfield[2]{#2}
\providecommand\bibinfo[2]{#2}
\providecommand\natexlab[1]{#1}
\providecommand\showeprint[2][]{arXiv:#2}

\bibitem[Babi{\'c} et~al\mbox{.}(2019)]%
        {babic2019fudge}
\bibfield{author}{\bibinfo{person}{Domagoj Babi{\'c}}, \bibinfo{person}{Stefan Bucur}, \bibinfo{person}{Yaohui Chen}, \bibinfo{person}{Franjo Ivan{\v{c}}i{\'c}}, \bibinfo{person}{Tim King}, \bibinfo{person}{Markus Kusano}, \bibinfo{person}{Caroline Lemieux}, \bibinfo{person}{L{\'a}szl{\'o} Szekeres}, {and} \bibinfo{person}{Wei Wang}.} \bibinfo{year}{2019}\natexlab{}.
\newblock \showarticletitle{Fudge: fuzz driver generation at scale}. In \bibinfo{booktitle}{\emph{Proceedings of the 2019 27th ACM Joint Meeting on European Software Engineering Conference and Symposium on the Foundations of Software Engineering}}. \bibinfo{pages}{975--985}.
\newblock


\bibitem[Chen et~al\mbox{.}(2023)]%
        {chen2023hopper}
\bibfield{author}{\bibinfo{person}{Peng Chen}, \bibinfo{person}{Yuxuan Xie}, \bibinfo{person}{Yunlong Lyu}, \bibinfo{person}{Yuxiao Wang}, {and} \bibinfo{person}{Hao Chen}.} \bibinfo{year}{2023}\natexlab{}.
\newblock \showarticletitle{Hopper: Interpretative fuzzing for libraries}. In \bibinfo{booktitle}{\emph{Proceedings of the 2023 ACM SIGSAC Conference on Computer and Communications Security}}. \bibinfo{pages}{1600--1614}.
\newblock


\bibitem[Deng et~al\mbox{.}(2023)]%
        {deng2023large}
\bibfield{author}{\bibinfo{person}{Yinlin Deng}, \bibinfo{person}{Chunqiu~Steven Xia}, \bibinfo{person}{Haoran Peng}, \bibinfo{person}{Chenyuan Yang}, {and} \bibinfo{person}{Lingming Zhang}.} \bibinfo{year}{2023}\natexlab{}.
\newblock \showarticletitle{Large language models are zero-shot fuzzers: Fuzzing deep-learning libraries via large language models}. In \bibinfo{booktitle}{\emph{Proceedings of the 32nd ACM SIGSOFT international symposium on software testing and analysis}}. \bibinfo{pages}{423--435}.
\newblock


\bibitem[Deng et~al\mbox{.}(2024)]%
        {10.1145/3597503.3623343}
\bibfield{author}{\bibinfo{person}{Yinlin Deng}, \bibinfo{person}{Chunqiu~Steven Xia}, \bibinfo{person}{Chenyuan Yang}, \bibinfo{person}{Shizhuo~Dylan Zhang}, \bibinfo{person}{Shujing Yang}, {and} \bibinfo{person}{Lingming Zhang}.} \bibinfo{year}{2024}\natexlab{}.
\newblock \showarticletitle{Large Language Models are Edge-Case Generators: Crafting Unusual Programs for Fuzzing Deep Learning Libraries}. In \bibinfo{booktitle}{\emph{Proceedings of the IEEE/ACM 46th International Conference on Software Engineering}} (Lisbon, Portugal) \emph{(\bibinfo{series}{ICSE '24})}. \bibinfo{publisher}{Association for Computing Machinery}, \bibinfo{address}{New York, NY, USA}, Article \bibinfo{articleno}{70}, \bibinfo{numpages}{13}~pages.
\newblock
\showISBNx{9798400702174}
\urldef\tempurl%
\url{https://doi.org/10.1145/3597503.3623343}
\showDOI{\tempurl}


\bibitem[Google(2023)]%
        {gtest}
\bibfield{author}{\bibinfo{person}{Google}.} \bibinfo{year}{2023}\natexlab{}.
\newblock \bibinfo{title}{{Google Test Framework}}.
\newblock
\newblock
\urldef\tempurl%
\url{https://github.com/google/googletest}
\showURL{%
\tempurl}


\bibitem[Green and Avgerinos(2022)]%
        {green2022graphfuzz}
\bibfield{author}{\bibinfo{person}{Harrison Green} {and} \bibinfo{person}{Thanassis Avgerinos}.} \bibinfo{year}{2022}\natexlab{}.
\newblock \showarticletitle{GraphFuzz: library API fuzzing with lifetime-aware dataflow graphs}. In \bibinfo{booktitle}{\emph{Proceedings of the 44th International Conference on Software Engineering}}. \bibinfo{pages}{1070--1081}.
\newblock


\bibitem[Hazan(2019)]%
        {DBLP:journals/corr/abs-1909-05207}
\bibfield{author}{\bibinfo{person}{Elad Hazan}.} \bibinfo{year}{2019}\natexlab{}.
\newblock \showarticletitle{Introduction to Online Convex Optimization}.
\newblock \bibinfo{journal}{\emph{CoRR}}  \bibinfo{volume}{abs/1909.05207} (\bibinfo{year}{2019}).
\newblock
\showeprint[arXiv]{1909.05207}
\urldef\tempurl%
\url{http://arxiv.org/abs/1909.05207}
\showURL{%
\tempurl}


\bibitem[Ispoglou et~al\mbox{.}(2020)]%
        {ispoglou2020fuzzgen}
\bibfield{author}{\bibinfo{person}{Kyriakos Ispoglou}, \bibinfo{person}{Daniel Austin}, \bibinfo{person}{Vishwath Mohan}, {and} \bibinfo{person}{Mathias Payer}.} \bibinfo{year}{2020}\natexlab{}.
\newblock \showarticletitle{$\{$FuzzGen$\}$: Automatic fuzzer generation}. In \bibinfo{booktitle}{\emph{29th USENIX Security Symposium (USENIX Security 20)}}. \bibinfo{pages}{2271--2287}.
\newblock


\bibitem[Jeong et~al\mbox{.}(2023)]%
        {jeong2023utopia}
\bibfield{author}{\bibinfo{person}{Bokdeuk Jeong}, \bibinfo{person}{Joonun Jang}, \bibinfo{person}{Hayoon Yi}, \bibinfo{person}{Jiin Moon}, \bibinfo{person}{Junsik Kim}, \bibinfo{person}{Intae Jeon}, \bibinfo{person}{Taesoo Kim}, \bibinfo{person}{WooChul Shim}, {and} \bibinfo{person}{Yong~Ho Hwang}.} \bibinfo{year}{2023}\natexlab{}.
\newblock \showarticletitle{Utopia: Automatic generation of fuzz driver using unit tests}. In \bibinfo{booktitle}{\emph{2023 IEEE Symposium on Security and Privacy (SP)}}. IEEE, \bibinfo{pages}{2676--2692}.
\newblock


\bibitem[Jiang et~al\mbox{.}(2021)]%
        {jiang2021rulf}
\bibfield{author}{\bibinfo{person}{Jianfeng Jiang}, \bibinfo{person}{Hui Xu}, {and} \bibinfo{person}{Yangfan Zhou}.} \bibinfo{year}{2021}\natexlab{}.
\newblock \showarticletitle{RULF: Rust library fuzzing via API dependency graph traversal}. In \bibinfo{booktitle}{\emph{2021 36th IEEE/ACM International Conference on Automated Software Engineering (ASE)}}. IEEE, \bibinfo{pages}{581--592}.
\newblock


\bibitem[Jung et~al\mbox{.}(2021)]%
        {jung2021winnie}
\bibfield{author}{\bibinfo{person}{Jinho Jung}, \bibinfo{person}{Stephen Tong}, \bibinfo{person}{Hong Hu}, \bibinfo{person}{Jungwon Lim}, \bibinfo{person}{Yonghwi Jin}, {and} \bibinfo{person}{Taesoo Kim}.} \bibinfo{year}{2021}\natexlab{}.
\newblock \showarticletitle{Winnie: Fuzzing windows applications with harness synthesis and fast cloning}. In \bibinfo{booktitle}{\emph{Proceedings of the 2021 Network and Distributed System Security Symposium (NDSS 2021)}}.
\newblock


\bibitem[Kang et~al\mbox{.}(2023a)]%
        {kang2023preliminary}
\bibfield{author}{\bibinfo{person}{Sungmin Kang}, \bibinfo{person}{Gabin An}, {and} \bibinfo{person}{Shin Yoo}.} \bibinfo{year}{2023}\natexlab{a}.
\newblock \showarticletitle{A preliminary evaluation of llm-based fault localization}.
\newblock \bibinfo{journal}{\emph{arXiv preprint arXiv:2308.05487}} (\bibinfo{year}{2023}).
\newblock


\bibitem[Kang et~al\mbox{.}(2023b)]%
        {kang2023explainable}
\bibfield{author}{\bibinfo{person}{Sungmin Kang}, \bibinfo{person}{Bei Chen}, \bibinfo{person}{Shin Yoo}, {and} \bibinfo{person}{Jian-Guang Lou}.} \bibinfo{year}{2023}\natexlab{b}.
\newblock \showarticletitle{Explainable automated debugging via large language model-driven scientific debugging}.
\newblock \bibinfo{journal}{\emph{arXiv preprint arXiv:2304.02195}} (\bibinfo{year}{2023}).
\newblock


\bibitem[Khoshmanesh and Lutz(2018)]%
        {khoshmanesh2018role}
\bibfield{author}{\bibinfo{person}{Seyedehzahra Khoshmanesh} {and} \bibinfo{person}{Robyn~R Lutz}.} \bibinfo{year}{2018}\natexlab{}.
\newblock \showarticletitle{The role of similarity in detecting feature interaction in software product lines}. In \bibinfo{booktitle}{\emph{2018 IEEE International Symposium on Software Reliability Engineering Workshops (ISSREW)}}. IEEE, \bibinfo{pages}{286--292}.
\newblock


\bibitem[Li et~al\mbox{.}(2024)]%
        {li2024acecoder}
\bibfield{author}{\bibinfo{person}{Jia Li}, \bibinfo{person}{Yunfei Zhao}, \bibinfo{person}{Yongmin Li}, \bibinfo{person}{Ge Li}, {and} \bibinfo{person}{Zhi Jin}.} \bibinfo{year}{2024}\natexlab{}.
\newblock \showarticletitle{AceCoder: An Effective Prompting Technique Specialized in Code Generation}.
\newblock \bibinfo{journal}{\emph{ACM Transactions on Software Engineering and Methodology}} (\bibinfo{year}{2024}).
\newblock


\bibitem[Li et~al\mbox{.}(2019)]%
        {li2019cerebro}
\bibfield{author}{\bibinfo{person}{Yuekang Li}, \bibinfo{person}{Yinxing Xue}, \bibinfo{person}{Hongxu Chen}, \bibinfo{person}{Xiuheng Wu}, \bibinfo{person}{Cen Zhang}, \bibinfo{person}{Xiaofei Xie}, \bibinfo{person}{Haijun Wang}, {and} \bibinfo{person}{Yang Liu}.} \bibinfo{year}{2019}\natexlab{}.
\newblock \showarticletitle{Cerebro: context-aware adaptive fuzzing for effective vulnerability detection}. In \bibinfo{booktitle}{\emph{Proceedings of the 2019 27th ACM Joint Meeting on European Software Engineering Conference and Symposium on the Foundations of Software Engineering}}. \bibinfo{pages}{533--544}.
\newblock


\bibitem[Lipowski and Lipowska(2012)]%
        {lipowski2012roulette}
\bibfield{author}{\bibinfo{person}{Adam Lipowski} {and} \bibinfo{person}{Dorota Lipowska}.} \bibinfo{year}{2012}\natexlab{}.
\newblock \showarticletitle{Roulette-wheel selection via stochastic acceptance}.
\newblock \bibinfo{journal}{\emph{Physica A: Statistical Mechanics and its Applications}} \bibinfo{volume}{391}, \bibinfo{number}{6} (\bibinfo{year}{2012}), \bibinfo{pages}{2193--2196}.
\newblock


\bibitem[Lyu et~al\mbox{.}(2024)]%
        {lyu2023prompt}
\bibfield{author}{\bibinfo{person}{Yunlong Lyu}, \bibinfo{person}{Yuxuan Xie}, \bibinfo{person}{Peng Chen}, {and} \bibinfo{person}{Hao Chen}.} \bibinfo{year}{2024}\natexlab{}.
\newblock \showarticletitle{Prompt Fuzzing for Fuzz Driver Generation}. In \bibinfo{booktitle}{\emph{Proceedings of the 2024 on {ACM} {SIGSAC} Conference on Computer and Communications Security, {CCS} 2024, Salt Lake City, UT, USA, October 14-18, 2024}}, \bibfield{editor}{\bibinfo{person}{Bo~Luo}, \bibinfo{person}{Xiaojing Liao}, \bibinfo{person}{Jun Xu}, \bibinfo{person}{Engin Kirda}, {and} \bibinfo{person}{David Lie}} (Eds.). \bibinfo{publisher}{{ACM}}, \bibinfo{pages}{3793--3807}.
\newblock
\urldef\tempurl%
\url{https://doi.org/10.1145/3658644.3670396}
\showDOI{\tempurl}


\bibitem[Meng et~al\mbox{.}(2024)]%
        {meng2024large}
\bibfield{author}{\bibinfo{person}{Ruijie Meng}, \bibinfo{person}{Martin Mirchev}, \bibinfo{person}{Marcel B{\"o}hme}, {and} \bibinfo{person}{Abhik Roychoudhury}.} \bibinfo{year}{2024}\natexlab{}.
\newblock \showarticletitle{Large language model guided protocol fuzzing}. In \bibinfo{booktitle}{\emph{Proceedings of the 31st Annual Network and Distributed System Security Symposium (NDSS)}}.
\newblock


\bibitem[Miao et~al\mbox{.}(2016)]%
        {miao2016fundamentals}
\bibfield{author}{\bibinfo{person}{Guowang Miao}, \bibinfo{person}{Jens Zander}, \bibinfo{person}{Ki~Won Sung}, {and} \bibinfo{person}{Slimane~Ben Slimane}.} \bibinfo{year}{2016}\natexlab{}.
\newblock \bibinfo{booktitle}{\emph{Fundamentals of mobile data networks}}.
\newblock \bibinfo{publisher}{Cambridge University Press}.
\newblock


\bibitem[OSS-Fuzz(2023)]%
        {ossfuzz}
\bibfield{author}{\bibinfo{person}{OSS-Fuzz}.} \bibinfo{year}{2023}\natexlab{}.
\newblock \bibinfo{title}{{Continuous fuzzing for open source software}}.
\newblock
\newblock
\urldef\tempurl%
\url{https://github.com/google/oss-fuzz}
\showURL{%
\tempurl}


\bibitem[Paria et~al\mbox{.}(2023)]%
        {paria2023divas}
\bibfield{author}{\bibinfo{person}{Sudipta Paria}, \bibinfo{person}{Aritra Dasgupta}, {and} \bibinfo{person}{Swarup Bhunia}.} \bibinfo{year}{2023}\natexlab{}.
\newblock \showarticletitle{Divas: An llm-based end-to-end framework for soc security analysis and policy-based protection}.
\newblock \bibinfo{journal}{\emph{arXiv preprint arXiv:2308.06932}} (\bibinfo{year}{2023}).
\newblock


\bibitem[Sch{\"a}fer et~al\mbox{.}(2023)]%
        {schafer2023empirical}
\bibfield{author}{\bibinfo{person}{Max Sch{\"a}fer}, \bibinfo{person}{Sarah Nadi}, \bibinfo{person}{Aryaz Eghbali}, {and} \bibinfo{person}{Frank Tip}.} \bibinfo{year}{2023}\natexlab{}.
\newblock \showarticletitle{An empirical evaluation of using large language models for automated unit test generation}.
\newblock \bibinfo{journal}{\emph{IEEE Transactions on Software Engineering}} (\bibinfo{year}{2023}).
\newblock


\bibitem[Snappy(2023)]%
        {snappy}
\bibfield{author}{\bibinfo{person}{Snappy}.} \bibinfo{year}{2023}\natexlab{}.
\newblock \bibinfo{title}{{A fast compressor/decompressor}}.
\newblock
\newblock
\urldef\tempurl%
\url{https://github.com/google/snappy}
\showURL{%
\tempurl}


\bibitem[Swi-prolog(2023a)]%
        {->}
\bibfield{author}{\bibinfo{person}{Swi-prolog}.} \bibinfo{year}{2023}\natexlab{a}.
\newblock \bibinfo{title}{{Predicate ->/2}}.
\newblock
\newblock
\urldef\tempurl%
\url{https://www.swi-prolog.org/pldoc/doc_for?object=(-%3E)/2}
\showURL{%
\tempurl}


\bibitem[Swi-prolog(2023b)]%
        {not}
\bibfield{author}{\bibinfo{person}{Swi-prolog}.} \bibinfo{year}{2023}\natexlab{b}.
\newblock \bibinfo{title}{{Predicate not}}.
\newblock
\newblock
\urldef\tempurl%
\url{https://www.swi-prolog.org/show-tag?tag=not}
\showURL{%
\tempurl}


\bibitem[Tonmoy et~al\mbox{.}(2024)]%
        {tonmoy2024comprehensive}
\bibfield{author}{\bibinfo{person}{SM Tonmoy}, \bibinfo{person}{SM Zaman}, \bibinfo{person}{Vinija Jain}, \bibinfo{person}{Anku Rani}, \bibinfo{person}{Vipula Rawte}, \bibinfo{person}{Aman Chadha}, {and} \bibinfo{person}{Amitava Das}.} \bibinfo{year}{2024}\natexlab{}.
\newblock \showarticletitle{A comprehensive survey of hallucination mitigation techniques in large language models}.
\newblock \bibinfo{journal}{\emph{arXiv preprint arXiv:2401.01313}} (\bibinfo{year}{2024}).
\newblock


\bibitem[Vikram et~al\mbox{.}(2023)]%
        {vikram2023can}
\bibfield{author}{\bibinfo{person}{Vasudev Vikram}, \bibinfo{person}{Caroline Lemieux}, {and} \bibinfo{person}{Rohan Padhye}.} \bibinfo{year}{2023}\natexlab{}.
\newblock \showarticletitle{Can large language models write good property-based tests?}
\newblock \bibinfo{journal}{\emph{arXiv preprint arXiv:2307.04346}} (\bibinfo{year}{2023}).
\newblock


\bibitem[Wielemaker et~al\mbox{.}(2010)]%
        {DBLP:journals/corr/abs-1011-5332}
\bibfield{author}{\bibinfo{person}{Jan Wielemaker}, \bibinfo{person}{Tom Schrijvers}, \bibinfo{person}{Markus Triska}, {and} \bibinfo{person}{Torbj{\"{o}}rn Lager}.} \bibinfo{year}{2010}\natexlab{}.
\newblock \showarticletitle{SWI-Prolog}.
\newblock \bibinfo{journal}{\emph{CoRR}}  \bibinfo{volume}{abs/1011.5332} (\bibinfo{year}{2010}).
\newblock
\showeprint[arXiv]{1011.5332}
\urldef\tempurl%
\url{http://arxiv.org/abs/1011.5332}
\showURL{%
\tempurl}


\bibitem[Xia et~al\mbox{.}(2024)]%
        {xia2024fuzz4all}
\bibfield{author}{\bibinfo{person}{Chunqiu~Steven Xia}, \bibinfo{person}{Matteo Paltenghi}, \bibinfo{person}{Jia Le~Tian}, \bibinfo{person}{Michael Pradel}, {and} \bibinfo{person}{Lingming Zhang}.} \bibinfo{year}{2024}\natexlab{}.
\newblock \showarticletitle{Fuzz4all: Universal fuzzing with large language models}. In \bibinfo{booktitle}{\emph{Proceedings of the IEEE/ACM 46th International Conference on Software Engineering}}. \bibinfo{pages}{1--13}.
\newblock


\bibitem[Xie et~al\mbox{.}(2023)]%
        {xie2023chatunitest}
\bibfield{author}{\bibinfo{person}{Zhuokui Xie}, \bibinfo{person}{Yinghao Chen}, \bibinfo{person}{Chen Zhi}, \bibinfo{person}{Shuiguang Deng}, {and} \bibinfo{person}{Jianwei Yin}.} \bibinfo{year}{2023}\natexlab{}.
\newblock \showarticletitle{ChatUniTest: a ChatGPT-based automated unit test generation tool}.
\newblock \bibinfo{journal}{\emph{arXiv preprint arXiv:2305.04764}} (\bibinfo{year}{2023}).
\newblock


\bibitem[Xu et~al\mbox{.}(2024)]%
        {xu2024code}
\bibfield{author}{\bibinfo{person}{Hanxiang Xu}, \bibinfo{person}{Wei Ma}, \bibinfo{person}{Ting Zhou}, \bibinfo{person}{Yanjie Zhao}, \bibinfo{person}{Kai Chen}, \bibinfo{person}{Qiang Hu}, \bibinfo{person}{Yang Liu}, {and} \bibinfo{person}{Haoyu Wang}.} \bibinfo{year}{2024}\natexlab{}.
\newblock \showarticletitle{A Code Knowledge Graph-Enhanced System for LLM-Based Fuzz Driver Generation}.
\newblock \bibinfo{journal}{\emph{arXiv preprint arXiv:2411.11532}} (\bibinfo{year}{2024}).
\newblock


\bibitem[Zhang et~al\mbox{.}(2023b)]%
        {10.5555/3620237.3620398}
\bibfield{author}{\bibinfo{person}{Cen Zhang}, \bibinfo{person}{Yuekang Li}, \bibinfo{person}{Hao Zhou}, \bibinfo{person}{Xiaohan Zhang}, \bibinfo{person}{Yaowen Zheng}, \bibinfo{person}{Xian Zhan}, \bibinfo{person}{Xiaofei Xie}, \bibinfo{person}{Xiapu Luo}, \bibinfo{person}{Xinghua Li}, \bibinfo{person}{Yang Liu}, {and} \bibinfo{person}{Sheikh~Mahbub Habib}.} \bibinfo{year}{2023}\natexlab{b}.
\newblock \showarticletitle{Automata-guided control-flow-sensitive fuzz driver generation}. In \bibinfo{booktitle}{\emph{Proceedings of the 32nd USENIX Conference on Security Symposium}} (Anaheim, CA, USA) \emph{(\bibinfo{series}{SEC '23})}. \bibinfo{publisher}{USENIX Association}, \bibinfo{address}{USA}, Article \bibinfo{articleno}{161}, \bibinfo{numpages}{18}~pages.
\newblock
\showISBNx{978-1-939133-37-3}


\bibitem[Zhang et~al\mbox{.}(2021)]%
        {zhang2021apicraft}
\bibfield{author}{\bibinfo{person}{Cen Zhang}, \bibinfo{person}{Xingwei Lin}, \bibinfo{person}{Yuekang Li}, \bibinfo{person}{Yinxing Xue}, \bibinfo{person}{Jundong Xie}, \bibinfo{person}{Hongxu Chen}, \bibinfo{person}{Xinlei Ying}, \bibinfo{person}{Jiashui Wang}, {and} \bibinfo{person}{Yang Liu}.} \bibinfo{year}{2021}\natexlab{}.
\newblock \showarticletitle{$\{$APICraft$\}$: Fuzz driver generation for closed-source $\{$SDK$\}$ libraries}. In \bibinfo{booktitle}{\emph{30th USENIX Security Symposium (USENIX Security 21)}}. \bibinfo{pages}{2811--2828}.
\newblock


\bibitem[Zhang et~al\mbox{.}(2024b)]%
        {10.1145/3650212.3680355}
\bibfield{author}{\bibinfo{person}{Cen Zhang}, \bibinfo{person}{Yaowen Zheng}, \bibinfo{person}{Mingqiang Bai}, \bibinfo{person}{Yeting Li}, \bibinfo{person}{Wei Ma}, \bibinfo{person}{Xiaofei Xie}, \bibinfo{person}{Yuekang Li}, \bibinfo{person}{Limin Sun}, {and} \bibinfo{person}{Yang Liu}.} \bibinfo{year}{2024}\natexlab{b}.
\newblock \showarticletitle{How Effective Are They? Exploring Large Language Model Based Fuzz Driver Generation}. In \bibinfo{booktitle}{\emph{Proceedings of the 33rd ACM SIGSOFT International Symposium on Software Testing and Analysis}} (Vienna, Austria) \emph{(\bibinfo{series}{ISSTA 2024})}. \bibinfo{publisher}{Association for Computing Machinery}, \bibinfo{address}{New York, NY, USA}, \bibinfo{pages}{1223–1235}.
\newblock
\showISBNx{9798400706127}
\urldef\tempurl%
\url{https://doi.org/10.1145/3650212.3680355}
\showDOI{\tempurl}


\bibitem[Zhang et~al\mbox{.}(2024a)]%
        {zhang2024llamafuzz}
\bibfield{author}{\bibinfo{person}{Hongxiang Zhang}, \bibinfo{person}{Yuyang Rong}, \bibinfo{person}{Yifeng He}, {and} \bibinfo{person}{Hao Chen}.} \bibinfo{year}{2024}\natexlab{a}.
\newblock \showarticletitle{LLAMAFUZZ: Large Language Model Enhanced Greybox Fuzzing}.
\newblock \bibinfo{journal}{\emph{arXiv preprint arXiv:2406.07714}} (\bibinfo{year}{2024}).
\newblock


\bibitem[Zhang et~al\mbox{.}(2023a)]%
        {zhang2023hallucination}
\bibfield{author}{\bibinfo{person}{Yue Zhang}, \bibinfo{person}{Yafu Li}, \bibinfo{person}{Leyang Cui}, \bibinfo{person}{Deng Cai}, \bibinfo{person}{Lemao Liu}, \bibinfo{person}{Tingchen Fu}, \bibinfo{person}{Xinting Huang}, \bibinfo{person}{Enbo Zhao}, \bibinfo{person}{Yu Zhang}, \bibinfo{person}{Yulong Chen}, \bibinfo{person}{Longyue Wang}, \bibinfo{person}{Anh~Tuan Luu}, \bibinfo{person}{Wei Bi}, \bibinfo{person}{Freda Shi}, {and} \bibinfo{person}{Shuming Shi}.} \bibinfo{year}{2023}\natexlab{a}.
\newblock \showarticletitle{Siren's Song in the AI Ocean: A Survey on Hallucination in Large Language Models}.
\newblock \bibinfo{journal}{\emph{arXiv preprint arXiv:2309.01219}} (\bibinfo{year}{2023}).
\newblock


\end{thebibliography}
